\documentclass[aps,twocolumn]{revtex4}

\usepackage{graphics}
\usepackage{epsfig}
\usepackage{bm}
\usepackage{amsmath,amssymb}

\def\ov#1{\overline{#1}}

\def\wt#1{\widetilde{#1}}
\def\vb#1{\mbox{\boldmath$#1$}}
\def\pd#1#2{\frac{\partial #1}{\partial #2}}
\def\fd#1#2{\frac{\delta #1}{\delta #2}}

\def\bdot{\,\vb{\cdot}\,}
\def\btimes{\,\vb{\times}\,}

\def\cal#1{\mathcal{#1}}

\def\exd{{\sf d}}

\newcommand{\bc}{\begin{center}}
\newcommand{\ec}{\end{center}}
\newcommand{\bt}{\begin{tabbing}}
\newcommand{\et}{\end{tabbing}}
\newcommand{\be}{\begin{equation}}
\newcommand{\ee}{\end{equation}}
\newcommand{\ben}{\begin{eqnarray}}
\newcommand{\een}{\end{eqnarray}}

\begin{document}

\title{Hamiltonian formulations for perturbed dissipationless plasma equations}

\author{A. J. Brizard$^{1}$ and C. Chandre$^2$}
\affiliation{$^1$Department of Physics, Saint Michael's College, Colchester, VT 05439, USA \\
$^2$CNRS/Aix-Marseille Universit\'{e}, I2M UMR7373, Avenue de Luminy, Case 907, 13009 Marseille, France}
    
\begin{abstract}
The Hamiltonian formulations for the perturbed Vlasov-Maxwell equations and the perturbed ideal magnetohydrodynamics (MHD) equations are expressed in terms of the perturbation derivative $\partial{\cal F}/\partial\epsilon \equiv [{\cal F}, {\cal S}]$ of an arbitrary functional ${\cal F}[\vb{\psi}]$ of the Vlasov-Maxwell fields $\vb{\psi} = ({\sf f},{\bf E},{\bf B})$ or the ideal MHD fields $\vb{\psi} = (\rho,{\bf u},s,{\bf B})$, which are assumed to depend continuously on the (dimensionless) perturbation parameter $\epsilon$. Here, $[\;,\;]$ denotes the functional Poisson bracket for each set of plasma equations and the perturbation {\it action} functional ${\cal S}$ is said to generate dynamically accessible perturbations of the plasma fields. The new Hamiltonian perturbation formulation introduces a framework for functional perturbation methods in plasma physics and highlights the crucial roles played by polarization and magnetization in Vlasov-Maxwell and ideal MHD perturbation theories. One application considered in this paper is a formulation of plasma stability that guarantees dynamical accessibility and leads to a natural generalization to higher-order perturbation theory.
\end{abstract}

\date{\today}


\maketitle

\section{Introduction}

The use of Hamiltonian perturbation methods in plasma physics \cite{Davidson,Cary_Kaufman_1981} has played a crucial role in our ability to understand the complex dynamics of collisionless magnetized plasmas. In particular, Lie-transform perturbation methods \cite{Dewar_1976,Kaufman_1978} have provided powerful pathways toward the dynamical reduction of the particle phase-space dynamics (e.g., the successive guiding-center \cite{Cary_Brizard_2009} and gyrocenter \cite{Brizard_Hahm_2007} reductions of charged-particle dynamics in strongly-magnetized plasmas), which are carried out as near-identity transformations that depend continuously on a small dimensionless ordering parameter $\epsilon$ \cite{Brizard_2001,Brizard_2008,Brizard_2009,Brizard_2018}. In general, applications of perturbation methods involve asymptotic expansions in powers of $\epsilon$, which are truncated at a predetermined maximum order. For example, the perturbation analyses of three-wave and four-wave interactions require truncations at third and fourth orders, respectively (for example, see the work of Boyd and Turner \cite{Boyd_Turner_1978} or the more recent work of Viscondi {\it et al.} \cite{Viscondi_2016}).

The Hamiltonian structures of the ideal magnetohydrodynamics (MHD) equations \cite{Morrison_Greene_1980} and the Vlasov-Maxwell equations \cite{M,PJM_1982,MW,B} have been known since the 1980s. Here, the field evolution equations $\partial_{t}\psi^{a} = [\psi^{a},\;{\cal H}]$ are expressed in Hamiltonian form in terms the Hamiltonian (energy) functional ${\cal H}$ and the functional Poisson bracket $[\;,\;]$, which satisfies the standard bracket properties. Hence, the evolution of an arbitrary functional ${\cal F}[\vb{\psi}]$ of the plasma dynamical fields $\vb{\psi}(x,t)$ is expressed as
\begin{equation}
\pd{\cal F}{t}[\vb{\psi}] \;=\; \left[{\cal F},\frac{}{} {\cal H}\right] \;\equiv\; \int \fd{\cal F}{\psi^{a}}\;\pd{\psi^{a}}{t}\;dx,
\label{eq:F_t}
\end{equation} 
where summation over repeated indices is implied throughout the paper and the integration domain may depend on the field-component $\psi^{a}$.

The purpose of the present paper is to introduce a Hamiltonian formulation of the perturbative evolution of the plasma dynamical fields $\vb{\psi}(x,t,\epsilon)$ parameterized by a continuous perturbation parameter $\epsilon$:
\begin{equation}
\pd{\cal F}{\epsilon}[\vb{\psi}]  \;=\; \left[{\cal F},\; {\cal S}\right] \;\equiv\; \int \fd{\cal F}{\psi^{a}}\;\pd{\psi^{a}}{\epsilon}\;dx,
\label{eq:F_epsilon}
\end{equation}
where the perturbation {\it action} functional ${\cal S}$ generates the plasma perturbations. This formulation was used by Hameiri \cite{Hameiri_2003} to investigate ideal MHD plasma stability (see Sec.~\ref{sec:Energy_pert}). In the present paper, it will be extended to include the Vlasov-Maxwell equations, and the connections between these two sets of dissipationless plasma equations will be explored.

Note that a common functional bracket $[\;,\;]$ is used in both Hamiltonian evolutions \eqref{eq:F_t}-\eqref{eq:F_epsilon} so that $\partial\psi^{a}/\partial t \equiv {\cal J}^{ab}\,\delta{\cal H}/\delta\psi^{b}$ and $\partial\psi^{a}/\partial \epsilon \equiv 
{\cal J}^{ab}\,\delta{\cal S}/\delta\psi^{b}$ are expressed in terms of the same anti-symmetric Poisson-matrix differential operator ${\cal J}^{ab}[\vb{\psi}]$. In addition, Casimir functionals ${\cal C}$ are naturally preserved under both dynamical and perturbative evolutions:
\begin{equation}
\left. \begin{array}{rcl}
\partial{\cal C}/\partial t & = & [{\cal C},\; {\cal H}] \;=\; 0 \\
 &&\\
\partial{\cal C}/\partial\epsilon & = & [{\cal C},\; {\cal S}] \;=\; 0
\end{array} \right\},
\label{eq:Casimir}
\end{equation}
since Casimir functionals satisfy the bracket identity $[{\cal C}, {\cal G}] \equiv 0$ for any functional ${\cal G}$. 
 
The standard linear perturbation theory is recovered from the functional perturbation equation \eqref{eq:F_epsilon}, with the definition of the Eulerian variation $\delta\psi^{a} \equiv (\partial\psi^{a}/\partial\epsilon)_{\epsilon = 0}$. The linear and nonlinear stability of kinetic (Vlasov) and fluid (magnetohydrodynamic) dissipationless plasma equilibria, on the other hand, are investigated through the second variation  $\delta^{2}{\cal F} \equiv \frac{1}{2}(\partial^{2} {\cal F}/\partial\epsilon^{2})_{\epsilon = 0}$ of the free-energy functional ${\cal F}$, where only dynamically-accessible perturbations \eqref{eq:F_epsilon} that preserve the Hamiltonian structure of the underlying plasma dynamics \cite{PJM_DP_1989,PJM_DP_1990,Brizard_1994,Hameiri_1998,Hameiri_2003,Andreussi_2013,Ilgisonis_Pastukhov_2000} are considered (see the review papers by Morrison \cite{PJM_1998,PJM_2005}, and references therein, as well as the mini-conference summary \cite{Brizard_ERT_2003}).

The remainder of the paper is organized as follows. In Secs.~\ref{sec:Ham_VM} and \ref{sec:Multiple}, we review the Hamiltonian formulation of the Vlasov-Maxwell equations and the multiple-time formulation of perturbed Hamiltonian dynamics \cite{Brizard_2001,Brizard_2018}, respectively. In Sec.~\ref{sec:Ham_pert_VM}, we use the Vlasov-Maxwell bracket structure introduced in Sec.~\ref{sec:Ham_VM} to formulate the Hamiltonian Vlasov-Maxwell perturbation theory based on Eq.~\eqref{eq:F_epsilon}. In this perturbation theory, the polarization and magnetization are naturally expressed in terms of Vlasov moments of the perturbed particle displacement $d{\bf x}/d\epsilon$. In Sec.~\ref{sec:Ham_pert_MHD}, as a second example of dissipationless plasma equations, we consider the Hamiltonian ideal MHD perturbation theory, where the ideal MHD bracket structure is now used in Eq.~\eqref{eq:F_epsilon}. Here, the interpretation of some of the perturbation functional derivatives 
$\delta{\cal S}/\delta\psi^{a}$ is presented through the generalized Clebsch representation of the fluid velocity ${\bf u}$. Next, we present a discussion of dynamical accessibility and plasma stability in Sec.~\ref{sec:Energy_pert} through the second-order perturbative derivative of the energy (Hamiltonian) functional for the Vlasov-Maxwell and ideal MHD equations. 

Lastly, we warn the reader about the notation used in the paper, where functions on particle phase space are denoted with sans-serif fonts while functionals of the Vlasov-Maxwell fields $({\sf f},{\bf E},{\bf B})$ are denoted with script fonts. For example, ${\sf H}$ denotes the single-particle Hamiltonian and ${\cal H}$ denotes the Hamiltonian functional, while ${\sf S}$ denotes the perturbation action (which perturbs orbits in particle phase space) and ${\cal S}$ denotes the perturbation action functional.

\section{\label{sec:Ham_VM}Hamiltonian formulation of the Vlasov-Maxwell equations}

We begin with a brief review the Hamiltonian formulation of the Vlasov-Maxwell equations, which presents a unifying principle for the self-consistent plasma interactions between charged particles and electromagnetic fields. 

First, the Hamiltonian dynamical evolution of the Vlasov distribution (along phase-space orbits for each particle species with mass $m$ and charge $e$) is governed by the Vlasov equation $d{\sf f}/dt = 0$, which is expressed as
\begin{eqnarray}
\pd{\sf f}{t} & = & -\;\{ {\sf f},\; {\sf H}\} \;+\; \frac{e}{c}\pd{\bf A}{t}\bdot\pd{\sf f}{\bf p} \nonumber \\
 & \equiv & -\;\{ {\sf f},\; {\sf K}\} \;-\; e\,{\bf E}\bdot\pd{\sf f}{\bf p},
\label{eq:Veq}
\end{eqnarray}
where ${\sf H} = e\,\Phi + {\sf K}$ is the sum of the electrostatic potential energy $e\,\Phi$ and the kinetic energy ${\sf K} = |{\bf p}|^{2}/2m$ (only non-relativistic results are shown in this paper) and the single-particle Poisson bracket is
\begin{equation}
\{ {\sf f},\; {\sf g}\} = \nabla f\bdot\pd{\sf g}{\bf p} - \pd{\sf f}{\bf p}\bdot\nabla{\sf g} \;+\; \frac{e}{c}\;{\bf B}\bdot\pd{\sf f}{\bf p}\btimes\pd{\sf g}{\bf p}.
\label{eq:xv_PB}
\end{equation}
The first two Maxwell equations (with particle sources) are
\begin{eqnarray}
\nabla\bdot{\bf E} & = & 4\pi\,\varrho \;\equiv\; 4\pi\,\int_{\bf p} e\,{\sf f}, \label{eq:E_div} \\
\nabla\btimes{\bf B} \;-\; \frac{1}{c}\pd{\bf E}{t} & = & \frac{4\pi}{c}\,{\bf J} \;\equiv\; \frac{4\pi}{c}\int_{\bf p} e\,{\sf f}\,{\bf v}, \label{eq:E_t}
\end{eqnarray}
where ${\bf v} = \partial {\sf K}/\partial{\bf p}$ denotes the particle's velocity and summation over particle species is included in the momentum-integral notation $\int_{\bf p} \equiv \sum \int d^{3}{\bf p}$. The remaining (source-free) Maxwell equations
\begin{eqnarray}
\partial{\bf B}/\partial t & = & -\;c\,\nabla\btimes{\bf E}, \label{eq:B_t} \\
\nabla\bdot{\bf B} & = & 0, \label{eq:div_B}
\end{eqnarray}
are Faraday's Law and the divergenceless condition for the magnetic field, respectively.
 
The Vlasov-Maxwell evolution equations \eqref{eq:Veq}, \eqref{eq:E_t}, and \eqref{eq:B_t} are expressed in Hamiltonian form \eqref{eq:F_t} using the Vlasov-Maxwell bracket~\cite{M,PJM_1982,MW,B}
\begin{eqnarray}
\left[{\cal F},\;{\cal G}\right] & = & \int_{\bf z}{\sf f} \left\{ \fd{{\cal F}}{\sf f},\; \fd{{\cal G}}{\sf f} \right\} \label{eq:MV_PB} \\
 & + & 4\pi \int_{\bf x} \fd{{\cal F}}{{\bf E}} \bdot \left( c\,\nabla\btimes\fd{{\cal G}}{{\bf B}} + \int_{\bf p} e\;\fd{{\cal G}}{\sf f}\pd{\sf f}{{\bf p}} \right)  \nonumber \\
 & - & 4\pi \int_{\bf x} \fd{{\cal G}}{{\bf E}} \bdot \left( c\,\nabla\btimes\fd{{\cal F}}{{\bf B}} + \int_{\bf p} e\;\fd{{\cal F}}{\sf f}\pd{\sf f}{{\bf p}} \right),
\nonumber
\end{eqnarray}
between two arbitrary functionals $\cal F$ and $\cal G$ of the Vlasov-Maxwell fields $({\sf f}, {\bf E}, {\bf B})$, and the Hamiltonian functional is
\begin{equation}
{\cal H}[{\sf f},{\bf E},{\bf B}] \;\equiv\; \int_{\bf z}{\sf K}\;{\sf f} \;+\; \int_{\bf x} \frac{1}{8\pi} \left( |{\bf E}|^{2} \;+\; |{\bf B}|^{2} \right),
\label{eq:H_def}
\end{equation}
which also corresponds to the energy invariant for the Vlasov-Maxwell equations. Here, we use the concise notation $\int_{\bf x} = \int d^{3}{\bf x}$ for a spatial integration over the field point ${\bf x}$ and the notation $\int_{\bf z} = \sum \int d^{3}{\bf x}\,d^{3}{\bf p}$ for a particle phase-space integration (including a summation over particle species). Using Eqs.~\eqref{eq:MV_PB}-\eqref{eq:H_def}, we thus obtain the Hamiltonian evolution functional equation \eqref{eq:F_t}:
\begin{eqnarray}
\pd{\cal F}{t} & = & -\;\int_{\bf z} \fd{{\cal F}}{\sf f}\left( \{{\sf f},\; {\sf K}\} + e\,{\bf E}\bdot\pd{\sf f}{\bf p}\right) \nonumber \\
 &  &+\; \int_{\bf x} \fd{{\cal F}}{{\bf E}} \bdot \left( c\;\nabla\btimes{\bf B} \;-\; 4\pi\,\int_{\bf p} e\;{\sf f}\,{\bf v}\right) \nonumber \\
 &  &-\; \int_{\bf x} \fd{{\cal F}}{{\bf B}} \bdot \left( c\frac{}{}\nabla\btimes{\bf E}\right) \label{eq:MV_PB_F} \\
  & \equiv & \int_{\bf z} \fd{\cal F}{\sf f}\pd{\sf f}{t} + \int_{\bf x} \left( \fd{\cal F}{\bf E}\bdot\pd{\bf E}{t} + \fd{\cal F}{\bf B}\bdot\pd{\bf B}{t} \right),
\nonumber
\end{eqnarray}
where we used the Vlasov-Maxwell equations \eqref{eq:Veq}, \eqref{eq:E_t}, and \eqref{eq:B_t}. We note that the Vlasov-Maxwell bracket \eqref{eq:MV_PB} satisfies the standard properties for a Poisson bracket, including the Jacobi identity (see Ref. \cite{PJM_1982} for details on a general proof and Refs.~\cite{PJM_2013,Brizard_2016_arxiv} for an explicit proof). In particular, the Jacobi property of the Vlasov-Maxwell bracket \eqref{eq:MV_PB} holds only if Eq.~\eqref{eq:div_B} is satisfied, which is a condition that is inherited from the Jacobi property of the single-particle Poisson bracket \eqref{eq:xv_PB}, which appears in Eq.~\eqref{eq:MV_PB}.

\section{\label{sec:Multiple}Multiple-time Hamiltonian dynamics}

It was previously shown (using canonical coordinates \cite{Brizard_2001,Brizard_2018}) that perturbed single-particle Hamiltonian dynamics of a charged particle (of mass $m$ and charge $e$) can be represented geometrically in terms of two Hamiltonian formulations. First, the Hamiltonian ${\sf H}$ acts as the generating function for infinitesimal phase-space transformations (parametrized by time $t$) and described by the noncanonical Hamilton equations 
\begin{eqnarray} 
\frac{dz^{\alpha}}{d t} & \equiv & \{ z^{\alpha},\; {\sf H}\} \;-\; \frac{e}{c}\pd{\bf A}{t}\bdot\{{\bf x}, z^{\alpha}\} \nonumber \\
 & = & \{ z^{\alpha},\; {\sf K} \} \;+\; e\,{\bf E}\bdot\{{\bf x}, z^{\alpha}\}.
\label{eq:Ham_h}
\end{eqnarray}
In addition, the electromagnetic potentials $(\Phi,{\bf A})$ are used to define the electromagnetic fields ${\bf E} = -\nabla\Phi - c^{-1}\partial{\bf A}/\partial t$ and ${\bf B} = \nabla\btimes{\bf A}$. 

Second, the action function ${\sf S}$ generates infinitesimal phase-space transformations (parametrized by the perturbation ordering parameter $\epsilon$), which are described by the noncanonical perturbation Hamilton equations
\begin{equation} 
\frac{dz^{\alpha}}{d\epsilon} \;\equiv\;  \left\{ z^{\alpha},\;{\sf S}\right\} \;-\; \frac{e}{c}\pd{\bf A}{\epsilon}\bdot\{ {\bf x},\; z^{\alpha}\}.
\label{eq:Ham_S}
\end{equation}
Here, the same single-particle Poisson bracket \eqref{eq:xv_PB} is used in both Hamilton evolution equations \eqref{eq:Ham_h}-\eqref{eq:Ham_S}. In addition, both Hamiltonian functions ${\sf H}$ and ${\sf S}$ depend on the noncanonical phase-space coordinates ${\bf z} = ({\bf x},{\bf p})$, the time $t$, and depend continuously on the perturbation parameter $\epsilon$ (with $\epsilon = 0$ representing an arbitrary time-dependent reference state). The electromagnetic potentials $(\Phi,{\bf A})$, on the other hand, depend on the space-time position $({\bf x},t)$ as well as the perturbation parameter $\epsilon$.

Each Hamilton equation \eqref{eq:Ham_h} and \eqref{eq:Ham_S} is derived by variational principle in App.~\ref{sec:App_A} and we note that each equation also satisfies its own Liouville theorem: 
$\partial(dz^{\alpha}/dt)/\partial z^{\alpha} \equiv 0 \equiv \partial(dz^{\alpha}/d\epsilon)/\partial z^{\alpha}$.

From a conceptual point of view, we assume that the order of temporal and perturbative evolutions is immaterial \cite{Brizard_2001,Brizard_2018}, i.e., evolving the reference state first ($t > 0$ at $\epsilon = 0$) and then perturbing it at a later time to a final perturbed state (with $\epsilon > 0$) should be equivalent to perturbing the reference state first ($\epsilon > 0$ at $t = 0$) and then evolving it to a final perturbed state (with $t > 0$). This assumption implies that the two noncanonical Hamiltonian dynamical $(d/dt)$ and perturbative $(d/d\epsilon)$ evolution operators 
\begin{eqnarray}
\frac{d}{dt} & \equiv & \pd{}{t} \;+\; \left\{\;,\frac{}{} {\sf H}\right\} \;-\; \frac{e}{c}\,\pd{\bf A}{t}\bdot\{{\bf x},\;\}, \label{eq:d_dt} \\
\frac{d}{d\epsilon} & \equiv & \pd{}{\epsilon} \;+\; \left\{\;,\frac{}{} {\sf S}\right\} \;-\; \frac{e}{c}\,\pd{\bf A}{\epsilon}\bdot\{{\bf x},\;\} \label{eq:d_de}
\end{eqnarray}
commute (see App.~\ref{sec:App_B} for a derivation)
\begin{eqnarray}
0 & = & \frac{d}{dt}\left(\frac{df}{d\epsilon}\right) \;-\; \frac{d}{d\epsilon}\left(\frac{df}{dt}\right) \nonumber \\
 & = & \left\{ f,\; \left(\frac{d{\sf S}}{dt} \;-\; \frac{d{\sf H}}{d\epsilon} \;-\; \left\{ {\sf S},\frac{}{} {\sf H}\right\} \right) \right\},
 \label{eq:constraint}
 \end{eqnarray}
where the function $f({\bf z},t,\epsilon)$ is arbitrary. Since this relation must hold for any function $f$, we obtain a constraint between the Hamiltonians ${\sf S}$ and ${\sf H}$:
\begin{equation}
\frac{d{\sf S}}{dt} \;=\; \frac{d{\sf H}}{d\epsilon} \;+\; \left\{ {\sf S},\frac{}{} {\sf H}\right\} \;\equiv\;  e\,\pd{\Phi}{\epsilon} \;-\; \frac{e}{c}\pd{\bf A}{\epsilon}\bdot{\bf v},
\label{eq:Sh_constraint}
\end{equation}
where ${\bf v} = \partial {\sf K}/\partial{\bf p} = {\bf p}/m$ denotes the particle's velocity. The constraint equation \eqref{eq:Sh_constraint}, which is also derived in App.~\ref{sec:App_A}, is a standard equation in Hamiltonian Lie-transform perturbation theory \cite{Dewar_1976,Kaufman_1978,Littlejohn_1982} and is a generalized form of the Hamilton-Jacobi equation that figures prominently in the perturbation analysis of the Vlasov equation \cite{Pfirsch_1966,Hori_1966,Kawakami_1970,Dewar_1973}. 

We note that the scalar field ${\sf S}$ in Eq.~\eqref{eq:Ham_S} generates canonical transformations of particle phase-space orbits, which in turn induce transformations on the Vlasov-Maxwell fields $({\sf f},{\bf E},{\bf B})$. In addition, we note that in order for Eqs.~\eqref{eq:Ham_S} and \eqref{eq:Sh_constraint} to be gauge-invariant, the function ${\sf S}$ is required to transform as
\begin{equation}
{\sf S} \;\rightarrow\; {\sf S} \;-\; \frac{e}{c}\;\pd{\chi}{\epsilon},
\label{eq:S_chi}
\end{equation}
where the gauge field $\chi({\bf x},t,\epsilon)$ generates the gauge transformation $(\Phi, {\bf A}) \rightarrow (\Phi - c^{-1}\partial\chi/\partial t,\; {\bf A} + \nabla\chi)$.

\section{\label{sec:Ham_pert_VM}Hamiltonian Vlasov-Maxwell Perturbation Theory}

We now turn our attention to perturbations of the Vlasov-Maxwell equations \eqref{eq:Veq}-\eqref{eq:div_B}. For this purpose, we introduce a second Hamiltonian formulation \eqref{eq:F_epsilon} of the Vlasov-Maxwell equations, where the perturbation action functional ${\cal S}[{\sf f},{\bf E},{\bf B}]$ generates the infinitesimal Vlasov-Maxwell perturbations of an arbitrary Vlasov-Maxwell functional ${\cal F}$ in terms of the perturbation parameter $\epsilon$. 

\subsection{Perturbed Vlasov equation}

The Hamiltonian perturbative evolution of the Vlasov distribution is governed by the perturbative Vlasov equation $d{\sf f}/d\epsilon = 0$, which is expressed as
\begin{equation}
\pd{\sf f}{\epsilon} \;=\; [{\sf f},\;{\cal S}] \;=\; -\;\left\{ {\sf f},\; \fd{\cal S}{\sf f}\right\} \;-\; 4\pi\,e\;\fd{\cal S}{\bf E}\bdot\pd{\sf f}{\bf p},
\label{eq:V_epsilon}
\end{equation}
where $(\delta{\cal S}/\delta {\sf f},\delta{\cal S}/\partial{\bf E})$ generate infinitesimal canonical transformations on particle phase space:
\begin{equation}
\frac{dz^{\alpha}}{d\epsilon} \;\equiv\; \left\{ z^{\alpha},\; \fd{\cal S}{\sf f}\right\} \;+\; 4\pi\,e\;\fd{\cal S}{\bf E}\bdot\pd{z^{\alpha}}{\bf p}.
\label{eq:V_depsilon}
\end{equation}
By comparing Eq.~\eqref{eq:V_depsilon} with Eq.~\eqref{eq:Ham_S}, we readily find the functional-derivative identities
\begin{equation}
\left( \fd{\cal S}{\sf f},\; \fd{\cal S}{\bf E}\right) \;\equiv\; \left( {\sf S},\; \frac{-1}{4\pi c}\pd{\bf A}{\epsilon} \right),
\label{eq:S_id}
\end{equation}
where we have omitted an arbitrary gauge function. Here, we note that the functional derivative $\delta{\cal S}/\delta{\bf B}$ is still unknown at this stage and, according to 
Eqs.~\eqref{eq:P_S}-\eqref{eq:M_S} derived below, it is involved in the invariance property associated with the definitions \eqref{eq:PM_gauge} of the polarization and magnetization for the Vlasov-Maxwell equations. 

\subsection{Perturbed charge and current densities}

The concept of dynamical accessibility of Vlasov perturbations \eqref{eq:V_epsilon} is associated with the fact that the phase-space integral of $\partial {\sf f}/\partial\epsilon$ vanishes at all orders of perturbation theory. Here, dynamical accessibility is extended to include the electromagnetic fields $({\bf E},{\bf B})$, whose Hamiltonian perturbative evolutions are expressed as
\begin{eqnarray}
\pd{\bf E}{\epsilon} & = & [{\bf E},\;{\cal S}] \;=\; 4\pi c\,\nabla\btimes\left(\fd{\cal S}{\bf B}\right) \;-\; 4\pi\,\mathbb{P}_{\epsilon}, \label{eq:E_epsilon} \\
\pd{\bf B}{\epsilon} & = & [{\bf B},\;{\cal S}] \;=\; \nabla\btimes\left(\pd{\bf A}{\epsilon}\right), \label{eq:B_epsilon}
\end{eqnarray}
where we have used the functional identities \eqref{eq:S_id} and we introduced the definition for the perturbation polarization
\begin{equation}
\mathbb{P}_{\epsilon} \equiv \int_{\bf p} e\,{\sf f}\;\frac{d{\bf x}}{d\epsilon} \;=\; \int_{\bf p} e\,{\sf f}\;\pd{\sf S}{\bf p},
\label{eq:P_epsilon}
\end{equation}
which is generated by the perturbation {\it displacement} $d{\bf x}/d\epsilon$. We note that the perturbation electromagnetic fields \eqref{eq:E_epsilon}-\eqref{eq:B_epsilon} satisfy the perturbed Maxwell equations
\begin{eqnarray}
0 & = & \nabla\bdot\left(\pd{\bf E}{\epsilon} \;+\; 4\pi\,\mathbb{P}_{\epsilon}\right), \label{eq:div_E_epsilon} \\
0 & = & \nabla\bdot\pd{\bf B}{\epsilon}, \label{eq:div_B_epsilon}
\end{eqnarray}
where Eq.~\eqref{eq:div_E_epsilon} follows from a perturbation of Gauss's Law \eqref{eq:E_div}:
\begin{equation}
\nabla\bdot\left(\pd{\bf E}{\epsilon}\right) = 4\pi\,\pd{\varrho}{\epsilon} \;=\; 4\pi\,[\varrho,\; {\cal S}] \;=\; -\;4\pi\,\nabla\bdot\mathbb{P}_{\epsilon},
 \label{eq:E_div_epsilon}
\end{equation}
while Eq.~\eqref{eq:div_B_epsilon} guarantees that magnetic perturbations $\partial{\bf B}/\partial\epsilon$ remain divergenceless.  In obtaining Eq.~\eqref{eq:E_div_epsilon}, we used the perturbation-derivative of the particle charge-density functional $\varrho({\bf r}) = \int_{\bf z} e\,\delta^{3}({\bf x} - {\bf r}){\sf f}$ according to Eq.~\eqref{eq:F_epsilon}.

Next, we take the perturbation-derivative of Maxwell's equation \eqref{eq:E_t}:
\begin{equation}
\nabla\btimes\left(\pd{\bf B}{\epsilon}\right) \;-\; \frac{1}{c}\pd{}{t}\left(\pd{\bf E}{\epsilon}\right) \;=\; \frac{4\pi}{c}\;\pd{\bf J}{\epsilon},
\label{eq:E_t_epsilon}
\end{equation}
where the perturbation-derivative of the particle current-density functional ${\bf J}({\bf r}) = \int_{\bf z} e{\bf v}\,\delta^{3}({\bf x} - {\bf r}){\sf f}$ is expressed according to Eq.~\eqref{eq:F_epsilon} as
\begin{eqnarray}
\pd{\bf J}{\epsilon} & = & [{\bf J},\; {\cal S}]\nonumber \\
  & = & \int_{\bf p} e\,{\sf f}\;\frac{d^{2}{\bf x}}{d\epsilon dt} \;-\; \nabla\bdot\left(\int_{\bf p}e\,{\sf f}\;\frac{d{\bf x}}{d\epsilon}\,\frac{d{\bf x}}{dt} \right),
\label{eq:J_epsilon_1}
\end{eqnarray}
with ${\bf v} \equiv d{\bf x}/dt$. Next, we use the perturbation polarization \eqref{eq:P_epsilon} to find the perturbation polarization current
\begin{eqnarray*}
\pd{\mathbb{P}_{\epsilon}}{t} & = & [\mathbb{P}_{\epsilon},\; {\cal H}] \\
 & = & \int_{\bf p} e\,{\sf f}\;\frac{d^{2}{\bf x}}{dt d\epsilon} \;-\; \nabla\bdot\left(\int_{\bf p}e\,{\sf f}\;\frac{d{\bf x}}{dt}\,\frac{d{\bf x}}{d\epsilon} \right).
\end{eqnarray*}
When this expression is inserted into Eq.~\eqref{eq:J_epsilon_1}, and using the symmetry $d(d{\bf x}/d\epsilon)/dt = d(d{\bf x}/dt)/d\epsilon$, which follows from the commutation of the Hamiltonian dynamical and perturbative flows based on Eq.~\eqref{eq:constraint},  we obtain the final expression for the perturbation-derivative of the particle current density
\begin{eqnarray}
\pd{\bf J}{\epsilon}  & = & \pd{\mathbb{P}_{\epsilon}}{t} \;+\; \nabla\bdot\left[\int_{\bf p} e\,{\sf f}\; \left( \frac{d{\bf x}}{dt}\,\frac{d{\bf x}}{d\epsilon} - \frac{d{\bf x}}{d\epsilon}\,\frac{d{\bf x}}{dt} \right) \right] \nonumber \\
 & \equiv & \pd{\mathbb{P}_{\epsilon}}{t} \;+\; c\;\nabla\btimes\mathbb{M}_{\epsilon},
 \label{eq:J_epsilon_final}
\end{eqnarray}
which is defined as the sum of the perturbed polarization and magnetization current densities.  Here, the perturbation magnetization current is expressed in terms of the perturbation magnetization
\begin{equation}
\mathbb{M}_{\epsilon} \equiv \int_{\bf p} \frac{e}{c}\,{\sf f} \left( \frac{d{\bf x}}{d\epsilon}\btimes\frac{d{\bf x}}{dt} \right) = \int_{\bf p} \frac{e}{c}\,{\sf f} \left( \pd{\sf S}{\bf p}\btimes\pd{H}{\bf p} \right),
\label{eq:M_epsilon}
\end{equation}
and Eq.~\eqref{eq:E_t_epsilon} becomes
\begin{equation}
\nabla\btimes\left(\pd{\bf B}{\epsilon}\right) - \frac{1}{c}\pd{}{t}\left(\pd{\bf E}{\epsilon}\right) = \frac{4\pi}{c}\pd{\mathbb{P}_{\epsilon}}{t} + 4\pi\nabla\btimes\mathbb{M}_{\epsilon},
\label{eq:Maxwell_epsilon}
\end{equation}
which may also be expressed as
\begin{eqnarray}
0 & = & \nabla\btimes\left(\pd{\bf B}{\epsilon} - 4\pi\,\mathbb{M}_{\epsilon}\right) - \frac{1}{c}\pd{}{t}\left(\pd{\bf E}{\epsilon} + 4\pi\,\mathbb{P}_{\epsilon}\right) \nonumber \\
 & \equiv & \nabla\btimes\left(\pd{\mathbb{H}}{\epsilon}\right) \;-\; \frac{1}{c}\,\pd{}{t}\left(\pd{\mathbb{D}}{\epsilon}\right),
\label{eq:Maxwell_macro}
\end{eqnarray}
where the macroscopic fields $\partial\mathbb{D}/\partial\epsilon$ and $\partial\mathbb{H}/\partial\epsilon$ are defined below.

The perturbative expressions for the polarization \eqref{eq:P_epsilon} and the magnetization \eqref{eq:M_epsilon} were first derived recently in Ref.~\cite{Brizard_2018} (see Sec.~VI) from a variational perturbation analysis of the Vlasov-Maxwell equations. These expressions are also expressed in terms of partial derivatives \eqref{eq:Lambda_EB} of the Lagrangian scalar field \eqref{eq:Lambda} with respect to the electromagnetic fields ${\bf E}$ and ${\bf B}$, respectively. 

We note that if we write ${\bf x} = {\bf x}_{0} + \epsilon\,\vb{\xi}_{1} + \cdots$, where $\vb{\xi}_{1}$ denotes the lowest-order particle displacement, the lowest-order contributions to the polarization \eqref{eq:P_epsilon} and the magnetization \eqref{eq:M_epsilon} are
\[ \left( \mathbb{P}_{\epsilon}, \frac{}{} \mathbb{M}_{\epsilon}\right) \simeq \int_{\bf p} e\,{\sf f} \left( \vb{\xi}_{1},\; \vb{\xi}_{1}\btimes\frac{1}{c}\frac{d{\bf x}_{0}}{dt} + \epsilon\;\vb{\xi}_{1}\btimes\frac{1}{c}\frac{d\vb{\xi}_{1}}{dt} \right), \]
where we retained the high-order intrinsic magnetic-dipole contribution $(e\,\epsilon\,\vb{\xi}_{1}\btimes c^{-1}d\vb{\xi}_{1}/dt)$ to the magnetization, in addition to the moving electric-dipole contribution $(e\,\vb{\xi}_{1}\btimes c^{-1}d{\bf x}_{0}/dt)$. Similar expressions were obtained by direct Lie-transform derivation for the general perturbed Vlasov-Maxwell equations \cite{Brizard_2008}, the reduced oscillation-center Vlasov-Maxwell equations \cite{Brizard_2009}, and the gyrocenter Vlasov-Maxwell equations \cite{Brizard_2013}. 

\subsection{Perturbed Maxwell equations}

The polarization equation \eqref{eq:E_epsilon} can also be used as a definition for the perturbation derivative of the displacement vector field $\mathbb{D}$:
\begin{equation}
\mathbb{P}_{\epsilon} = -\frac{1}{4\pi}\pd{\bf E}{\epsilon} + c\,\nabla\vb{\times}\left(\fd{\cal S}{\bf B}\right) \equiv -\,\frac{1}{4\pi} \pd{}{\epsilon}\left({\bf E} - \mathbb{D}\right).
\label{eq:P_S}
\end{equation}
If we substitute Eq.~\eqref{eq:E_epsilon} into Eq.~\eqref{eq:Maxwell_epsilon}, we obtain the magnetization equation
\begin{equation}
\mathbb{M}_{\epsilon} = \frac{1}{4\pi}\;\pd{\bf B}{\epsilon} - \pd{}{t}\left(\fd{\cal S}{\bf B}\right) \equiv  \frac{1}{4\pi}\pd{}{\epsilon}({\bf B} - \mathbb{H}),
\label{eq:M_S}
\end{equation}
which defines the perturbation derivative of the vector field $\mathbb{H}$. Here, the definitions of the macroscopic fields $(\mathbb{D},\mathbb{H})$ in terms of the functional derivative $\delta{\cal S}/\delta{\bf B}$ guarantee that $\nabla\bdot(\partial\mathbb{D}/\partial\epsilon) \equiv 0$ and that Eq.~\eqref{eq:Maxwell_macro} is satisfied identically.

We note that Eqs.~\eqref{eq:P_S}-\eqref{eq:M_S} involve the standard invariance property of the polarization and magnetization \cite{Healy_1978,Goedecke_1998}, where the transformation
\begin{equation}
\left. \begin{array}{rcl}
\mathbb{P}^{\prime} & = & \mathbb{P} \;-\; c\,\nabla\btimes\vb{\Lambda} \\
\mathbb{M}^{\prime} & = & \mathbb{M} \;+\; \partial\vb{\Lambda}/\partial t
\end{array} \right\}
\label{eq:PM_gauge}
\end{equation} 
yields identical polarization charge densities and polarization and magnetization current densities:
\begin{equation}
\left. \begin{array}{rcl}
\nabla\bdot\mathbb{P}^{\prime} & = & \nabla\bdot\mathbb{P} \\
\partial\mathbb{P}^{\prime}/\partial t + c\nabla\btimes\mathbb{M}^{\prime} & = & \partial\mathbb{P}/\partial t + c\nabla\btimes\mathbb{M}
\end{array} \right\},
\label{eq:PM_invariant}
\end{equation}  
where $\vb{\Lambda}$ is an arbitrary vector field. Hence, the specific expression for the functional derivative $\delta{\cal S}/\delta{\bf B}$ has no impact on the polarization charge density or the polarization-magnetization current density.

\subsection{Vlasov-Maxwell Perturbation Theory}

Now that we have established the perturbative evolution of the Vlasov-Maxwell equations, we can now explore, first, the linearized Vlasov-Maxwell equations and, second, the fully-perturbed equations.

\subsubsection{Linearized Vlasov-Maxwell equations}

As an application of the perturbative evolution of the Vlasov-Maxwell equations, we consider the simple example of the linearized Vlasov-Maxwell equations obtained in the absence of background electric and magnetic fields \cite{Kaufman_Cohen_2019}. Here, using the Fourier space-time decomposition of the first-order fields
\[ ({\sf f}_{1},{\bf E}_{1},{\bf B}_{1}) \;=\; {\rm Re}\left[\left(\wt{\sf f}_{1}, \wt{\bf E}_{1}, \wt{\bf B}_{1}\right)\; e^{i({\bf k}\bdot{\bf x} - \omega\,t)}\right], \]
the linearized Vlasov-Maxwell equations are derived from Eqs.~\eqref{eq:Veq}-\eqref{eq:E_t} and expressed as
\begin{eqnarray}
-i\,\omega^{\prime}\;\wt{\sf f}_{1} & = & -e\left(\wt{\bf E}_{1} + \frac{\bf v}{c}\btimes\wt{\bf B}_{1}\right)\vb{\cdot}\pd{{\sf f}_{0}}{\bf p}, \label{eq:f1_eq} \\
i{\bf k}\bdot\wt{\bf E}_{1} & = & 4\pi\,\int_{\bf p}e\,\wt{\sf f}_{1} \;\equiv\; 4\pi\,\wt{\varrho}_{1}, \label{eq:k.E1} \\
i{\bf k}\btimes\wt{\bf B}_{1} + i\,\frac{\omega}{c}\,\wt{\bf E}_{1} & = & \frac{4\pi}{c}\,\int_{\bf p}e{\bf v}\,\wt{\sf f}_{1} \;\equiv\; \frac{4\pi}{c}\,\wt{\bf J}_{1} \label{eq:kxB1},
\end{eqnarray}
where the time-independent and uniform background Vlasov distribution ${\sf f}_{0}$ is chosen to be consistent with vanishing fields $({\bf E}_{0},{\bf B}_{0}) = (0,0)$, while the Doppler-shifted frequency is $\omega^{\prime} \equiv \omega - {\bf k}\bdot{\bf v}$.

First, we turn our attention to the first-order equation $\partial {\sf S}_{1}/\partial t + {\bf v}\bdot\nabla {\sf S}_{1} = e\,(\Phi_{1} - {\bf A}_{1}\bdot{\bf v}/c)$ derived from Eq.~\eqref{eq:Sh_constraint}, which is Fourier-decomposed as
\begin{equation}
\wt{\sf S}_{1} \;=\; \frac{i\,e}{\omega^{\prime}} \left(\wt{\Phi}_{1} \;-\; \frac{\bf v}{c}\bdot\wt{\bf A}_{1}\right).
\end{equation}
From this expression, we calculate the first-order perturbed displacement from Eq.~\eqref{eq:Ham_S}:
\begin{equation}
\wt{\vb{\xi}}_{1} \;\equiv\; \pd{\wt{\sf S}_{1}}{\bf p} \;=\; -\,\frac{e}{m\,\omega^{\prime 2}}\left(\wt{\bf E}_{1} + \frac{\bf v}{c}\btimes\wt{\bf B}_{1}\right),
\label{eq:xi_1}
\end{equation}
where $\wt{\bf E}_{1} = -\,i{\bf k}\wt{\Phi}_{1} + i\wt{\bf A}_{1}\,\omega/c$ and $\wt{\bf B}_{1} = i{\bf k}\btimes\wt{\bf A}_{1}$. From Eq.~\eqref{eq:xi_1}, we can now define the first-order polarization and magnetization from Eqs.~\eqref{eq:P_epsilon} and
\eqref{eq:M_epsilon}:
\begin{equation}
\left(\wt{\mathbb{P}}_{1},\; \wt{\mathbb{M}}_{1}\right) \equiv \int_{\bf p} e\,{\sf f}_{0} \left( \wt{\vb{\xi}}_{1},\; \wt{\vb{\xi}}_{1}\btimes\frac{\bf v}{c}\right).
\label{eq:PM_1}
\end{equation}
We note that Eq.~\eqref{eq:PM_1} can be derived from the second-order ponderomotive Hamiltonian functional \cite{Cary_Kaufman_1981} 
\begin{equation}
\ov{\cal K}_{2} \;=\; \int_{\bf p}{\sf f}_{0}\,m\,\omega^{\prime 2}\,|\wt{\vb{\xi}}_{1}|^{2}
\end{equation}
associated with these first-order field perturbations, which yields the standard expressions $\wt{\mathbb{P}}_{1} \equiv -\,\delta\ov{\cal K}_{2}/\delta\wt{\bf E}_{1}^{*}$ and $\wt{\mathbb{M}}_{1} \equiv -\,\delta\ov{\cal K}_{2}/\delta\wt{\bf B}_{1}^{*}$.

Next, we define the perturbed particle momentum from Eq.~\eqref{eq:Ham_S}:
\begin{equation}
\wt{\bf p}_{1} \;\equiv\; -\,i{\bf k}\,\wt{\sf S}_{1} - (e/c)\,\wt{\bf A}_{1} \;=\; -i\,m\omega^{\prime}\,\wt{\vb{\xi}}_{1},
\end{equation}
which turns Eq.~\eqref{eq:f1_eq} into $\wt{\sf f}_{1} = -\,\wt{\bf p}_{1}\bdot\partial {\sf f}_{0}/\partial{\bf p}$ (where we assumed that $\nabla {\sf f}_{0} \equiv 0$). Using the definition \eqref{eq:PM_1} for the first-order polarization, the perturbed charge density becomes
\begin{eqnarray}
\wt{\varrho}_{1} & = &\int_{\bf p}e\,\wt{\sf f}_{1} \;=\; \int_{\bf p}e\,{\sf f}_{0}\;\left(\pd{}{\bf p}\bdot\wt{\bf p}_{1}\right) \nonumber \\
 & = & \int_{\bf p} e\,\left(-i{\bf k}\bdot\wt{\vb{\xi}}_{1}\right)\; {\sf f}_{0} \;\equiv\; -\,i\,{\bf k}\bdot\wt{\mathbb{P}}_{1},
\end{eqnarray}
after integration by parts, so that Eq.~\eqref{eq:div_E_epsilon} [and Eq.~\eqref{eq:k.E1}] is now expressed as 
\begin{equation}
i{\bf k}\bdot\left(\wt{\bf E}_{1} \;+\; 4\pi\,\wt{\mathbb{P}}_{1}\right) \;\equiv\; i{\bf k}\bdot\wt{\mathbb{D}}_{1}\;=\; 0. 
\label{eq:kD_1} 
\end{equation}
Upon integration by parts, on the other hand, the perturbed current density becomes
\begin{eqnarray}
\wt{\bf J}_{1} & = & -\,i \int_{\bf p}e^{2}\frac{\bf v}{\omega^{\prime}} \left(\wt{\bf E}_{1} + \frac{\bf v}{c}\btimes\wt{\bf B}_{1}\right)\bdot\pd{{\sf f}_{0}}{\bf p} \nonumber \\
 & = & -\,i \int_{\bf p} e\,{\sf f}_{0}\left[ \omega^{\prime}\,\wt{\vb{\xi}}_{1} \;+\;  {\bf v} \left({\bf k}\bdot\wt{\vb{\xi}}_{1}\right) \right] \nonumber \\
  & \equiv & -i\,\omega\;\wt{\mathbb{P}}_{1} \;+\; i\,{\bf k}c\btimes\wt{\mathbb{M}}_{1},
 \end{eqnarray}
 where we introduced the definitions \eqref{eq:PM_1} for the first-order polarization and magnetization. Hence, Eq.~\eqref{eq:Maxwell_epsilon} [and Eq.~\eqref{eq:kxB1}] is now expressed as 
 \begin{equation}
 i\,{\bf k}c\btimes\wt{\mathbb{H}}_{1}\;+\; i\,\omega\; \wt{\mathbb{D}}_{1} \;=\; 0,
 \label{eq:kxH_1}
 \end{equation}
 where
 \begin{equation}
 \left. \begin{array}{rcl}
\wt{\bf E}_{1} + 4\pi\,\wt{\mathbb{P}}_{1} & = & i\,{\bf k}c\btimes\wt{\vb{\cal R}}_{1} \;\equiv\; \wt{\mathbb{D}}_{1} \\
\wt{\bf B}_{1} - 4\pi\,\wt{\mathbb{M}}_{1} & = & -i\,\omega\,\wt{\vb{\cal R}}_{1}  \;\equiv\; \wt{\mathbb{H}}_{1}
\end{array} \right\},
\end{equation}
which are consistent with Eqs.~\eqref{eq:kD_1} and \eqref{eq:kxH_1}, where $\wt{\vb{\cal R}}_{1}$ is related to the first-order term in the functional derivative $\delta{\cal S}/\delta{\bf B}$.

Lastly, we note that the eikonal-averaged second-order polarization  $\ov{\mathbb{P}}_{2} \equiv \int_{\bf p}{\sf f}_{0}\,\ov{\vb{\pi}}_{2}$ and magnetization $\ov{\mathbb{M}}_{2} \equiv \int_{\bf p}{\sf f}_{0}\,(\ov{\vb{\mu}}_{2} + \ov{\vb{\pi}}_{2}\btimes{\bf v}/c)$ were expressed \cite{Cary_Kaufman_1981,Brizard_2009} in terms of the ponderomotive electric and magnetic dipole moments 
\begin{equation}
\left( \begin{array}{c}
\ov{\vb{\pi}}_{2} \\
\ov{\vb{\mu}}_{2} \end{array} \right) \;\equiv\; \left( \begin{array}{c}
e\,{\bf k}\btimes(i\,\wt{\vb{\xi}}_{1}\btimes\wt{\vb{\xi}}_{1}^{*})\\
(e/c)\,\omega^{\prime}(i\,\wt{\vb{\xi}}_{1}\btimes\wt{\vb{\xi}}_{1}^{*}).
\end{array} \right) 
\end{equation}
These second-order expressions satisfy the equations ${\bf k}\bdot\ov{\mathbb{P}}_{2} = 0$ and $c{\bf k}\btimes\ov{\mathbb{M}}_{2} - \omega\,\ov{\mathbb{P}}_{2} = 0$.

\subsubsection{Functional perturbation theory}

We are now able to write an explicit expression that connects the reference Vlasov-Maxwell fields $({\sf f}_{0},{\bf E}_{0},{\bf B}_{0})$ to the perturbed Vlasov-Maxwell fields $({\sf f},{\bf E},{\bf B})$ by using the relation
\begin{equation}
({\sf f},{\bf E},{\bf B}) \equiv ({\sf f}_{0},{\bf E}_{0},{\bf B}_{0}) + \int_{0}^{\epsilon} \left(\pd{\sf f}{\sigma}, \pd{\bf E}{\sigma}, \pd{\bf B}{\sigma}\right)\;d\sigma.
\end{equation}
By integrating the charge and current perturbation derivatives $(\partial\varrho/\partial\sigma,\partial{\bf J}/\partial\sigma)$, we find the general relations for particle charge and current densities
\begin{eqnarray}
\varrho & \equiv & \varrho_{0} \;-\; \nabla\bdot\mathbb{P}, \label{eq:rho_P} \\
{\bf J} & \equiv & {\bf J}_{0} \;+\; \pd{\mathbb{P}}{t} \;+\; c\,\nabla\btimes\mathbb{M}, \label{eq:J_PM} 
\end{eqnarray}
where the total polarization and magnetization
\begin{eqnarray}
(\mathbb{P}, \mathbb{M}) & \equiv & \int_{0}^{\epsilon}(\mathbb{P}_{\sigma}, \mathbb{M}_{\sigma})\;d\sigma \\
 & \simeq & \int_{\bf p} e\,{\sf f} \left( \epsilon\,\vb{\xi}_{1},\; \epsilon\,\vb{\xi}_{1}\btimes\frac{1}{c}\frac{d{\bf x}_{0}}{dt} + \frac{\epsilon^{2}}{2}\;\vb{\xi}_{1}\btimes\frac{1}{c}\frac{d\vb{\xi}_{1}}{dt} \right) \nonumber
\end{eqnarray} 
are expressed in terms of standard $\epsilon$ expansions \cite{Brizard_2008,Brizard_2009,Brizard_2013}. Hence, all perturbations of the charge and current densities \eqref{eq:rho_P}-\eqref{eq:J_PM} are expressed solely in terms of polarization and magnetization effects. This conclusion was also reached through a perturbation expansion of the Vlasov-Maxwell equations \cite{Brizard_2018}.

Next, we integrate Eqs.~\eqref{eq:E_epsilon} and \eqref{eq:M_S} to obtain the electromagnetic relations
\begin{eqnarray}
{\bf E} + 4\pi\,\mathbb{P} & \equiv & \mathbb{D} \;=\; {\bf E}_{0} \;+\; c\,\nabla\btimes{\bf K}, 
\label{eq:EPD} \\
{\bf B} - 4\pi\,\mathbb{M} &\equiv& \mathbb{H} \;=\; {\bf B}_{0} \;+\; \pd{\bf K}{t},
\label{eq:BMH}
\end{eqnarray}
where the gauge vector field ${\bf K}$ is
\[ {\bf K} \;\equiv\; 4\pi\int_{0}^{\epsilon}\fd{\cal S}{\bf B}\;d\sigma. \]
We readily verify that
\begin{equation}
\nabla\bdot \mathbb{D} \;=\; \nabla\bdot{\bf E}_{0} \;\equiv\; 4\pi\,\varrho_{0}.
\end{equation}
and
\begin{equation}
\nabla\btimes \mathbb{H} - \frac{1}{c}\pd{\mathbb{D}}{t} = \nabla\btimes{\bf B}_{0} - \frac{1}{c}\pd{{\bf E}_{0}}{t} \;\equiv\; \frac{4\pi}{c}\,{\bf J}_{0}.
\label{eq:curl_H}
\end{equation}
We note, here, the vector field ${\bf K}$ appearing in Eqs.~\eqref{eq:EPD} and \eqref{eq:BMH} cancels out in Eq.~\eqref{eq:curl_H}.
 
 \section{\label{sec:Ham_pert_MHD}Hamiltonian Formulation of Perturbed Ideal Magnetohydrodynamics}

As a second example of a set of dissipationless plasma equations with a Hamiltonian structure, we explore the Hamiltonian formulation of ideal magnetohydrodynamics (MHD) \cite{Morrison_Greene_1980}, which has the bracket structure
\begin{widetext}
\begin{eqnarray}
\left[{\cal F},\; {\cal G}\right] &=& \int_{\bf x}\left[ \rho^{-1}\nabla\btimes{\bf u}\bdot\left(\fd{\cal F}{\bf u}\btimes\fd{\cal G}{\bf u}\right) \;-\; \left( \fd{\cal F}{\rho}\,\nabla\bdot\fd{\cal G}{\bf u} + \fd{\cal F}{\bf u}\bdot\nabla\fd{\cal G}{\rho}\right) \right] \nonumber \\
 &&+\; \int_{\bf x} \left[ \rho^{-1} \nabla s\bdot\left(\fd{\cal F}{\bf u}\,\fd{\cal G}{s} - \fd{\cal F}{s}\,\fd{\cal G}{\bf u}\right) \;-\; \rho^{-1}\fd{\cal F}{\bf u}\bdot{\bf B}\btimes\nabla\btimes\fd{\cal G}{\bf B} - \fd{\cal F}{\bf B}\bdot\nabla\btimes\left({\bf B}\btimes\rho^{-1}
 \fd{\cal G}{\bf u}\right) \right],
 \label{eq:MHD}
\end{eqnarray}
\end{widetext}
with functionals of the mass density $\rho$, the plasma single-fluid velocity ${\bf u}$, the entropy per unit mass $s$, and the magnetic field ${\bf B}$ (which is assumed to be divergenceless \cite{Morrison_Greene_1980}). Using the Hamiltonian functional
\begin{equation}
{\cal H} \;=\; \int_{\bf x} \left(\frac{\rho}{2}\,|{\bf u}|^{2} \;+\; \rho\,U(\rho,s) \;+\; \frac{1}{8\pi}\,|{\bf B}|^{2}\right),
\label{eq:Ham_MHD}
\end{equation}
where the internal energy density (per unit mass) $U(\rho,s)$ satisfies the First Law of Thermodynamics $dU = T\,ds + (P/\rho^{2})\,d\rho$, the equations of ideal MHD are expressed in Hamiltonian form $\partial_{t}\psi^{a} = [\psi^{a},{\cal H}]$ as
\begin{eqnarray}
\partial_{t}\rho &=& -\,\nabla\bdot(\rho\,{\bf u}), \label{eq:MHD_rho} \\
\partial_{t}{\bf u} &=& -\,{\bf u}\bdot\nabla{\bf u} + \rho^{-1}\left({\bf J}\btimes{\bf B}/c \;-\frac{}{} \nabla P\right), \label{eq:MHD_u} \\
\partial_{t}s &=& -\;{\bf u}\bdot\nabla s, \label{eq:MHD_s} \\
\partial_{t}{\bf B} &=& \nabla\btimes({\bf u}\btimes{\bf B}), \label{eq:MHD_B}
\end{eqnarray}
where ${\bf J} = (c/4\pi)\,\nabla\btimes{\bf B}$ denotes the plasma current density.

\subsection{Dynamically-accessible perturbed ideal MHD}

We now use the ideal MHD bracket \eqref{eq:MHD} to evaluate the perturbative derivatives $\partial_{\epsilon}\psi^{a} = [\psi^{a},{\cal S}]$: 
\begin{eqnarray}
\partial_{\epsilon}\rho &=& -\,\nabla\bdot\fd{\cal S}{\bf u}, \label{eq:delta_rho} \\
\partial_{\epsilon}{\bf u} &=& -\;\nabla\left(\fd{\cal S}{\rho}\right) \;+\; \rho^{-1}\fd{\cal S}{\bf u}\btimes\nabla\btimes{\bf u} \nonumber \\
 &&+\; \rho^{-1}\left( \nabla s\;\fd{\cal S}{s} \;-\; {\bf B}\btimes\nabla\btimes\fd{\cal S}{\bf B} \right), \label{eq:delta_u} \\
\partial_{\epsilon}s &=& -\;\rho^{-1}\nabla s\bdot\fd{\cal S}{\bf u}, \label{eq:delta_s} \\
\partial_{\epsilon}{\bf B} &=& -\;\nabla\btimes\left({\bf B}\btimes\rho^{-1}\fd{\cal S}{\bf u}\right), \label{eq:delta_B}
\end{eqnarray}
in terms of the perturbation action functional ${\cal S}$. We note that these expressions have been derived previously \cite{Hameiri_1998,Hameiri_2003,Andreussi_2013,Kaltsas_2020} within the concept of dynamical accessibility of allowed perturbations of ideal MHD equilibria with flows. If we introduce the ideal MHD fluid displacement 
\begin{equation}
\vb{\xi} \;\equiv\; \rho^{-1}\delta{\cal S}/\delta{\bf u}, 
\label{eq:xi_def}
\end{equation}
we recover the standard expressions \cite{Newcomb_1962}
\begin{equation}
\left. \begin{array}{rcl}
\partial_{\epsilon}\rho &=& -\,\nabla\bdot\left(\rho\frac{}{}\vb{\xi}\right) \\ 
\partial_{\epsilon}s &=& -\,\vb{\xi}\bdot\nabla s \\
\partial_{\epsilon}{\bf B} &=& \nabla\btimes\left(\vb{\xi}\btimes\frac{}{}{\bf B}\right) 
\end{array} \right\}
\label{eq:const_rhoBs}
\end{equation}
from Eqs.~\eqref{eq:delta_rho}, \eqref{eq:delta_s}, and \eqref{eq:delta_B}, respectively. 

We now identify ${\bf x}$ as the position of a fluid element, so that the fluid velocity is defined as ${\bf u} \equiv d{\bf x}/dt$ while the fluid displacement is defined $\vb{\xi} \equiv d{\bf x}/d\epsilon$. Hence, using the commutation relations of the operators 
$d/dt = \partial_{t} + {\bf u}\bdot\nabla$ and $d/d\epsilon = \partial_{\epsilon} + \vb{\xi}\bdot\nabla$, the identity $d{\bf u}/d\epsilon = d\vb{\xi}/dt$ yields the standard expression for the perturbative derivative of the fluid velocity \cite{Newcomb_1962}
\begin{equation}
\partial_{\epsilon}{\bf u} \;=\; \partial_{t}\vb{\xi} \;+\; {\bf u}\bdot\nabla\vb{\xi} \;-\; \vb{\xi}\bdot\nabla{\bf u}. 
\label{eq:const_u}
\end{equation}
Equation \eqref{eq:delta_u}, on the other hand, becomes 
\begin{eqnarray}
\partial_{\epsilon}{\bf u} &=& -\;\nabla\left(\fd{\cal S}{\rho}\right) \;+\; \vb{\xi}\btimes\nabla\btimes{\bf u} \nonumber \\
 &&+\; \rho^{-1}\left( \nabla s\;\fd{\cal S}{s} \;-\; {\bf B}\btimes\nabla\btimes\fd{\cal S}{\bf B} \right),
 \label{eq:u_epsilon}
 \end{eqnarray}
 whose expression is reminiscent of the Clebsch representation for the fluid velocity ${\bf u}$ derived from a Lagrangian variational formulation of ideal MHD (see also Errata cited in Ref.~\cite{Morrison_Greene_1980}). Here, the functional derivatives $(\delta{\cal S}/\delta\rho, \delta{\cal S}/\delta s, \delta{\cal S}/\delta{\bf B})$ represent additional degrees of freedom \cite{Hameiri_1998,Hameiri_2003,Andreussi_2013} for dynamically accessible perturbations of the fluid velocity. We can also obtain an expression for the total time derivative of the fluid displacement
 \[ \rho\,\frac{d\vb{\xi}}{dt} = \nabla\psi^{a}\fd{\cal S}{\psi^{a}} + \nabla\vb{\cdot}\left[ {\bf B}\,\fd{\cal S}{\bf B} - {\bf I} \left( \rho\,\fd{\cal S}{\rho} + {\bf B}\bdot\fd{\cal S}{\bf B}\right)\right], \]
  which is obtained by comparing Eqs.~\eqref{eq:const_u}-\eqref{eq:u_epsilon}. 
   
 \subsection{Lagrangian variational principle for ideal MHD}
 
 In this Section, we review the Lagrangian variational derivation of the ideal MHD equations \eqref{eq:MHD_rho}-\eqref{eq:MHD_B}. By far the simplest Lagrangian derivation is based on the constrained variational principle \cite{Newcomb_1962} based on the Lagrangian density $L = \frac{1}{2}\,\rho|{\bf u}|^{2} - \rho\,U(\rho,s) - |{\bf B}|^{2}/8\pi$ with the constrained variations $(\delta\rho,\delta{\bf u},\delta s, \delta{\bf B})$ expressed in terms of Eqs.~\eqref{eq:const_rhoBs}-\eqref{eq:const_u}, with the Eulerian variation $\delta(\;) \equiv \partial_{\epsilon}(\;)|_{\epsilon = 0}$. While Eqs.~\eqref{eq:MHD_rho}, \eqref{eq:MHD_s}-\eqref{eq:MHD_B} are immediately recovered from Eqs.~\eqref{eq:const_rhoBs} through the substitutions $(\partial_{\epsilon} \rightarrow \partial_{t},\; \vb{\xi} \rightarrow {\bf u})$, Eq.~\eqref{eq:MHD_u} is obtained from the Euler-Lagrange equation with respect to $\vb{\xi}$.
 
 Many variational principles for ideal MHD are expressed in terms of Clebsch variables to represent the fluid velocity ${\bf u}$ and the magnetic field ${\bf B}$. Some of the earliest examples include the works of Calkin \cite{Calkin_1963}, Seliger and Whitham \cite{Seliger_1968}, Merches \cite{Merches_1969}, and Nassar and Putterman \cite{Nassar_1985}, while historical surveys are included in Morrison's review paper \cite{PJM_2005} and the recent works of Yahalom \cite{Yahalom_2016,Yahalom_Qin_2020}.  The most relevant work for our present purpose, however, is presented by Merches \cite{Merches_1969}, whose notation is modified here to match our own notation. 
 
 We thus begin with the Lagrangian density \cite{Merches_1969,footnote}
 \begin{eqnarray}
 L &=& \frac{1}{2}\,\rho\,|{\bf u}|^{2} \;-\; \rho\,U(\rho,s) \;-\; \frac{1}{8\pi}\left(|{\bf E}|^{2} \;-\frac{}{} |{\bf B}|^{2} \right) \nonumber \\
  &&+\; \alpha\,\left( \frac{d\rho}{dt} + \rho\,\nabla\bdot{\bf u}\right) \;-\; \rho\,\beta\;\frac{ds}{dt},
  \label{eq:Lag_MHD}
  \end{eqnarray}
  where the Lagrange multipliers $\alpha$ and $\beta$ are used to enforce the conservation laws of mass and entropy, respectively (with $d/dt = \partial/\partial t + {\bf u}\bdot\nabla$). In Eq.~\eqref{eq:Lag_MHD}, Merches \cite{Merches_1969} introduces the following representations for the electric and magnetic fields (also used by Calkin \cite{Calkin_1963})
  \begin{equation}
  \left. \begin{array}{rcl}
  {\bf E} &=& {\bf E}_{0} \;-\; 4\pi\,\mathbb{P} \;+\; c\,\nabla\btimes{\bf K} \\
  {\bf B} &=& {\bf B}_{0} \;+\; 4\pi\,\mathbb{P}\btimes{\bf u}/c \;+\; \partial{\bf K}/\partial t
  \end{array} \right\},
  \label{eq:EB_MHD}
  \end{equation}
  where we have added the reference fields $({\bf E}_{0}, {\bf B}_{0})$ in order to use the same Vlasov-Maxwell expressions \eqref{eq:EPD}-\eqref{eq:BMH} derived from the Vlasov-Maxwell bracket, with the magnetization derived from the Vlasov-Maxwell expression \eqref{eq:M_epsilon} as $\mathbb{M} = \mathbb{P}\btimes{\bf u}/c$ (i.e., it only includes the moving electric-dipole contribution). Here, the polarization $\mathbb{P}$ and the gauge vector field ${\bf K}$ are both independent variational fields, in addition to the ideal MHD fields $(\rho,{\bf u},s)$ and the Lagrange multipliers $(\alpha,\beta)$. 
  
The charge density $\varrho$ is derived from Gauss's Law: $\nabla\bdot{\bf E} = 4\pi\varrho$, where $\varrho = \varrho_{0} - \nabla\bdot\mathbb{P}$. The current density ${\bf J}$, on the other hand, is derived from the Maxwell equation
$c\,\nabla\btimes{\bf B} - \partial{\bf E}/\partial t = 4\pi\,{\bf J}$, where ${\bf J} = {\bf J}_{0} + \nabla\btimes(\mathbb{P}\btimes{\bf u}) + \partial\mathbb{P}/\partial t$. In Ref.~\cite{Panofsky_Phillips_1955}, the charge and current densities $(\varrho_{0},
{\bf J}_{0})$ are identified as {\it true} densities, and based on this analysis, the current density is defined by Calkin  \cite{Calkin_1963} and Merches \cite{Merches_1969} as
\begin{equation}
{\bf J} \;\equiv\;  \partial\mathbb{P}/\partial t \;+\; \nabla\btimes(\mathbb{P}\btimes{\bf u}) \;+\; {\bf u}\,(\nabla\bdot\mathbb{P}),
\label{eq:J_Merches}
\end{equation}
with contributions from the polarization and magnetization current densities, respectively, as well as the convective polarization-charge current density ${\bf u}\,(\nabla\bdot\mathbb{P})$ \cite{Panofsky_Phillips_1955,Roa-Neri_2018}. 
  
The Euler-Lagrange equations 
\[ \pd{}{t}\left(\pd{L}{(\partial_{t}\chi^{a})} \right) \;+\; \nabla\bdot\left(\pd{L}{(\nabla\chi^{a})}\right) \;=\; \pd{L}{\chi^{a}}, \]
for $\chi^{a} = (\alpha,\beta,\rho,{\bf u},s,\mathbb{P},{\bf K})$ are now, respectively, expressed as Eqs.~\eqref{eq:MHD_rho} and \eqref{eq:MHD_s}, and
\begin{eqnarray}
\partial_{t}\alpha + \nabla\bdot(\alpha\,{\bf u}) &=& \alpha\,(\nabla\bdot{\bf u}) + \frac{1}{2}\,|{\bf u}|^{2} \nonumber \\
 &&-\;(U + P/\rho), \label{eq:rho_var} \\
\nabla(\rho\,\alpha) &=& \rho{\bf u} + \alpha\,\nabla\rho - \rho\beta\,\nabla s \nonumber \\
 &&+\; {\bf B}\btimes\mathbb{P}/c,\label{eq:u_var}  \\
\partial_{t}(\rho\beta) \;+\; \nabla\bdot(\rho\beta\,{\bf u}) &=& \rho\,T, \label{eq:s_var} \\
0&=& {\bf E} \;+\; {\bf u}\btimes{\bf B}/c, \label{eq:P_var} \\
\partial_{t}{\bf B} \;+\; c\,\nabla\btimes{\bf E} &=& 0. \label{eq:K_var}
\end{eqnarray}
It is simple to see how Eqs.~\eqref{eq:P_var}-\eqref{eq:K_var} are combined to give Eq.~\eqref{eq:MHD_B}. The remaining equations \eqref{eq:rho_var}-\eqref{eq:s_var} are now expressed as
\begin{equation}
\left. \begin{array}{rcl}
d\alpha/dt &=& |{\bf u}|^{2}/2 \;-\; (U + P/\rho) \\
d\beta/dt &=& T
\end{array} \right\},
\label{eq:alpha_beta}
\end{equation}
with the generalized Clebsch representation for the fluid velocity
\begin{equation}
{\bf u} \;=\; \nabla\alpha + \beta\,\nabla s + \mathbb{P}\btimes{\bf B}/(\rho c),
\label{eq:u_Clebsch}
\end{equation}
where polarization drives a cross-field flow.  We note that the equations \eqref{eq:alpha_beta} can also be written in terms of the Hamiltonian functional \eqref{eq:Ham_MHD} as
 \begin{eqnarray}
 \frac{d\alpha}{dt} &\equiv& \frac{1}{\rho}\,\fd{\cal H}{\bf u}\bdot{\bf u} \;-\; \fd{\cal H}{\rho}, \label{eq:alpha_t_H} \\
 \frac{d\beta}{dt} &\equiv& \frac{1}{\rho}\;\fd{\cal H}{s}, \label{eq:beta_t_H} 
 \end{eqnarray}
 while Eq.~\eqref{eq:J_Merches} can be expressed as
 \begin{equation}
\pd{\mathbb{P}}{t} \;+\; \nabla\btimes(\mathbb{P}\btimes{\bf u}) \;+\; {\bf u}\,(\nabla\bdot\mathbb{P}) \;\equiv\; c\,\nabla\btimes\fd{\cal H}{\bf B}.
\label{eq:J_t_H}
\end{equation}
We note that the Lagrangian variational principle of Nassar and Putterman \cite{Nassar_1985} obtain the fluid-velocity expression ${\bf u} = \nabla\alpha + \beta\,\nabla s + (\nabla\btimes{\bf K})\btimes{\bf B}/\rho$, instead of Eq.~\eqref{eq:u_Clebsch}.

We now reconstruct Eq.~\eqref{eq:MHD_u} from Eqs.~\eqref{eq:alpha_beta}-\eqref{eq:u_Clebsch} as follows. First, we begin with taking the gradient of the first equation in Eq.~\eqref{eq:alpha_beta}:
\[ \partial_{t}\nabla\alpha \;+\; \nabla\left({\bf u}\bdot\nabla\alpha\right) \;=\; \nabla{\bf u}\bdot{\bf u} \;-\; \rho^{-1}\nabla P \;-\; T\,\nabla s, \]
where we used $\nabla U = T\,\nabla s + (P/\rho^{2})\,\nabla\rho$. Second, we use Eqs.~\eqref{eq:alpha_beta}-\eqref{eq:u_Clebsch} to find
\begin{eqnarray*}
\partial_{t}\nabla\alpha &=& \partial_{t}{\bf u} \;-\; T\,\nabla s \;+\; \nabla({\bf u}\bdot\beta\nabla s)  \\
 &&-\; {\bf u}\btimes\left(\nabla\beta \btimes\nabla s\right) \;-\; \partial_{t}\left( \mathbb{P}\btimes\frac{}{}{\bf B}/(\rho c)\right),
 \end{eqnarray*}
 so that, after cancellations, we obtain
 \begin{eqnarray}
 \partial_{t}{\bf u} &=& \nabla{\bf u}\bdot{\bf u} - \nabla\left[{\bf u}\bdot\frac{}{}(\nabla\alpha + \beta\,\nabla s)\right] + {\bf u}\btimes\left(\nabla\beta \btimes\nabla s\right) \nonumber \\
  &&-\; \rho^{-1}\,\nabla P \;+\; \partial_{t}\left( \mathbb{P}\btimes\frac{}{}{\bf B}/(\rho c)\right).
  \label{eq:partial_u}
  \end{eqnarray}
  Third, using Eqs.~\eqref{eq:u_Clebsch} and the vorticity expression
\begin{equation} 
\nabla\btimes{\bf u} = \nabla\beta\btimes\nabla s + \nabla\btimes\left(\mathbb{P}\btimes\frac{}{}{\bf B}/(\rho c)\right),
\label{eq:vor_MHD}
\end{equation}
we obtain the expressions
  \begin{eqnarray*}
 \nabla\left[{\bf u}\bdot\frac{}{}(\nabla\alpha + \beta\,\nabla s)\right] &=& \nabla|{\bf u}|^{2} - \nabla\left[{\bf u}\bdot\left( \mathbb{P}\btimes\frac{}{}{\bf B}/(\rho c)\right)\right] \\
 {\bf u}\btimes\left(\nabla\beta \btimes\nabla s\right) &=& \nabla{\bf u}\bdot{\bf u} \;-\; {\bf u}\bdot\nabla{\bf u} \\
  &&-\; {\bf u}\btimes\nabla\btimes\left( \mathbb{P}\btimes\frac{}{}{\bf B}/(\rho c)\right),
 \end{eqnarray*}
 so that Eq.~\eqref{eq:partial_u} becomes
 \begin{eqnarray*}
 \frac{d{\bf u}}{dt} + \rho^{-1}\nabla P &=& \partial_{t}\left( \mathbb{P}\btimes\frac{}{}{\bf B}/(\rho c)\right) \\
  &&-\; {\bf u}\btimes\nabla\btimes\left( \mathbb{P}\btimes\frac{}{}{\bf B}/(\rho c)\right) \\
   &&+\; \nabla\left[{\bf u}\bdot\left( \mathbb{P}\btimes\frac{}{}{\bf B}/(\rho c)\right) \right] \\
   &=& \left[ \pd{\mathbb{P}}{t} + \nabla\btimes(\mathbb{P}\btimes{\bf u}) + {\bf u}\,(\nabla\bdot\mathbb{P}) \right]\btimes\frac{\bf B}{\rho c},
 \end{eqnarray*}
 after using the ideal MHD equations \eqref{eq:MHD_rho} and \eqref{eq:MHD_B}, as well as the vector identity
 \begin{eqnarray} 
 \nabla({\bf A}\bdot{\bf B}\btimes{\bf C}) &=& {\bf A}\btimes\nabla\btimes({\bf B}\btimes{\bf C}) + {\bf B}\btimes\nabla\btimes({\bf C}\btimes{\bf A}) \nonumber \\
  &&+\; {\bf C}\btimes\nabla\btimes({\bf A}\btimes{\bf B}) - ({\bf A}\btimes{\bf B}) \nabla\bdot{\bf C} \nonumber \\
   &&-\; ({\bf B}\btimes{\bf C}) \nabla\bdot{\bf A} - ({\bf C}\btimes{\bf A}) \nabla\bdot{\bf B},
   \label{eq:vector_id}
 \end{eqnarray}
 which holds for any three vector fields $({\bf A},{\bf B},{\bf C})$. Lastly, after using the definition \eqref{eq:J_Merches} for the current density, we recover Eq.~\eqref{eq:MHD_u}.
 
 Hence, we have shown that the ideal MHD Lagrangian density \eqref{eq:Lag_MHD}, with the Merches-Calkin representation \eqref{eq:EB_MHD} for the electric and magnetic fields, yields the ideal MHD equations \eqref{eq:MHD_rho}-\eqref{eq:MHD_B}.

\subsection{Perturbed fluid velocity}

In concluding this Section, we would now like to reconcile the perturbative derivative of the fluid velocity \eqref{eq:u_epsilon} with the perturbative derivative of Eq.~\eqref{eq:u_Clebsch}:
\begin{widetext}
\begin{eqnarray}
\pd{\bf u}{\epsilon} &=& \nabla\left(\frac{d\alpha}{d\epsilon} - \vb{\xi}\bdot\nabla\alpha\right) \;+\; \left(\frac{d\beta}{d\epsilon} - \vb{\xi}\bdot\nabla\beta\right)\,\nabla s \;-\; \nabla\left(\vb{\xi}\bdot\beta\,\nabla s\right) \;+\; (\vb{\xi}\bdot\nabla s)\,\nabla\beta \;+\; \pd{\mathbb{P}}{\epsilon}\btimes\frac{\bf B}{\rho c} \nonumber \\
  &&+ \frac{\mathbb{P}}{\rho c}\btimes\nabla\btimes(\vb{\xi}\btimes{\bf B}) + \nabla\bdot(\rho\,\vb{\xi})\;\mathbb{P}\btimes\frac{\bf B}{\rho^{2}c}, \nonumber \\
   &=& \nabla\left(\frac{d\alpha}{d\epsilon} \;-\; \vb{\xi}\bdot{\bf u}\right) \;+\;  \frac{d\beta}{d\epsilon}\;\nabla s \;+\; \vb{\xi}\btimes\nabla\btimes{\bf u} \;+\; \nabla\left[\vb{\xi}\bdot\left(\mathbb{P}\btimes\frac{}{}{\bf B}/(\rho c)\right) \right] \;-\; \vb{\xi}\btimes\nabla\btimes\left(\mathbb{P}\btimes\frac{}{}{\bf B}/(\rho c)\right) \nonumber \\
    &&+\; \left[ \pd{\mathbb{P}}{\epsilon} \;+\; \nabla\bdot(\rho\,\vb{\xi})\;\frac{\mathbb{P}}{\rho}\right]\btimes\frac{\bf B}{\rho c} + \frac{\mathbb{P}}{\rho c}\btimes\nabla\btimes(\vb{\xi}\btimes{\bf B}),
\label{eq:epsilon_u}
 \end{eqnarray}
\end{widetext}
where we inserted the definition $d/d\epsilon = \partial/\partial\epsilon + \vb{\xi}\bdot\nabla$ and we used the perturbation derivatives \eqref{eq:const_rhoBs}. 

Next, using again the vector identity \eqref{eq:vector_id}, Eq.~\eqref{eq:epsilon_u} becomes
\begin{eqnarray}
\pd{\bf u}{\epsilon} &=& \nabla\left(\frac{d\alpha}{d\epsilon} \;-\; \vb{\xi}\bdot{\bf u}\right) \;+\;  \frac{d\beta}{d\epsilon}\;\nabla s \;+\; \vb{\xi}\btimes\nabla\btimes{\bf u} \nonumber \\
 &&+\; \left[ \pd{\mathbb{P}}{\epsilon} + \nabla\btimes(\mathbb{P}\btimes\vb{\xi}) + \vb{\xi}\;(\nabla\bdot\mathbb{P}) \right]\btimes\frac{\bf B}{\rho c}.
 \label{eq:epsilon_u_final}
 \end{eqnarray}
 When compared with Eq.~\eqref{eq:u_epsilon}, we obtain the Clebsch-variable perturbation derivatives
 \begin{eqnarray}
 \frac{d\alpha}{d\epsilon} &\equiv& \frac{1}{\rho}\,\fd{\cal S}{\bf u}\bdot{\bf u} \;-\; \fd{\cal S}{\rho}, \label{eq:alpha_epsilon_S} \\
 \frac{d\beta}{d\epsilon} &\equiv& \frac{1}{\rho}\;\fd{\cal S}{s}, \label{eq:beta_epsilon_S}
 \end{eqnarray}
while the polarization perturbation derivative is 
 \begin{equation}
 \pd{\mathbb{P}}{\epsilon} + \nabla\btimes(\mathbb{P}\btimes\vb{\xi}) + \vb{\xi}\;(\nabla\bdot\mathbb{P}) \;\equiv\; c\,\nabla\btimes\fd{\cal S}{\bf B}. \label{eq:J_epsilon_S}
 \end{equation}
 This last equation implies that the polarization charge density $q \equiv -\,\nabla\bdot\mathbb{P}$ satisfies the standard charge conservation law: $\partial q/\partial\epsilon + \nabla\bdot(q\,\vb{\xi}) = 0$. Lastly, the similarities between Eqs.~\eqref{eq:alpha_t_H}-\eqref{eq:J_t_H} and Eqs.~\eqref{eq:alpha_epsilon_S}-\eqref{eq:J_epsilon_S} show a common Hamiltonian structure.
 
 \section{\label{sec:Energy_pert}Hamiltonian perturbation of dynamical plasma invariants}
 
 While Casimir invariants are naturally preserved by the Hamiltonian perturbation framework considered here, we now explore how the energy-momentum functionals of the Vlasov-Maxwell equations
 \begin{eqnarray}
{\cal H} &=&\int_{\bf z} {\sf f}\,{\sf K} \;+\; \frac{1}{8\pi}\int_{\bf x} \left( |{\bf E}|^{2} \;+\frac{}{} |{\bf B}|^{2} \right),
\label{eq:energy_MV} \\
\vb{\cal P} &=& \int_{\bf z} {\sf f}\,{\bf p} \;+\; \int_{\bf x} \frac{{\bf E}\btimes{\bf B}}{4\pi c},
\label{eq:momentum_MV}
\end{eqnarray}
and the ideal MHD equations
\begin{eqnarray}
{\cal H} &=& \int_{\bf x} \left( \frac{1}{2}\,\rho\,|{\bf u}|^{2} \;+\; \rho\,U(\rho,s) \;+\; \frac{1}{8\pi}\,|{\bf B}|^{2} \right),
\label{eq:energy_MHD} \\
\vb{\cal P} &=&  \int_{\bf x} \rho\,{\bf u},
\label{eq:momentum_MHD}
\end{eqnarray}
are perturbed within our formulation based on Eq.~\eqref{eq:F_epsilon}.
 
 \subsection{Perturbed Vlasov-Maxwell energy-momentum}
 
The perturbation derivative \eqref{eq:F_epsilon} of the Vlasov-Maxwell energy functional is
\begin{equation}
\pd{\cal H}{\epsilon} = \int_{\bf z}{\sf f}\,\frac{d{\sf K}}{d\epsilon} + \int_{\bf x}\left(\frac{\bf E}{4\pi}\vb{\cdot}\pd{\bf E}{\epsilon} + \frac{\bf B}{4\pi}\vb{\cdot}\pd{\bf B}{\epsilon}\right),
\label{eq:Energy_epsilon}
\end{equation}
where
\begin{eqnarray} 
\frac{d{\sf K}}{d\epsilon} &=& \{ {\sf K},\; {\sf S}\} \;-\; \frac{e}{c}\,{\bf v}\bdot\pd{\bf A}{\epsilon} \nonumber \\
   &=& \left(\pd{\sf S}{t} - e\,\pd{\Phi}{\epsilon}\right) \;+\;  e\,{\bf E}\bdot\pd{\sf S}{\bf p}.
\end{eqnarray}
\begin{equation}
\pd{\cal H}{\epsilon} = \int_{\bf z}{\sf f}\, \left(\pd{\sf S}{t} - e\,\pd{\Phi}{\epsilon}\right) + \int_{\bf x}\left(\frac{\bf E}{4\pi}\vb{\cdot}\pd{\mathbb{D}}{\epsilon} + \frac{\bf B}{4\pi}\vb{\cdot}\pd{\bf B}{\epsilon}\right).
\label{eq:Energy_final}
\end{equation}
We note that $\partial{\cal H}/\partial\epsilon \equiv -\,\partial{\cal S}/\partial t = -\,[{\cal S}, {\cal H}]$ because of the antisymmetry of the bracket $[\;,\;]$. Hence, the Hamiltonian perturbation
\[ \pd{\cal H}{\epsilon} \;=\; -\,\int_{\bf z}\pd{\sf f}{t}\;{\sf S} - \int_{\bf x} \left(\pd{\bf E}{t}\bdot\fd{\cal S}{\bf E} + \pd{\bf B}{t}\bdot\fd{\cal S}{\bf B}\right) \]
vanishes for Vlasov-Maxwell equilibria (i.e., when $\partial/\partial t \equiv 0$).

The perturbation derivative \eqref{eq:F_epsilon} of the Vlasov-Maxwell momentum functional is
\begin{equation}
\pd{\vb{\cal P}}{\epsilon} = \int_{\bf z}{\sf f}\,\frac{d{\bf p}}{d\epsilon} + \int_{\bf x}\left(\pd{\bf E}{\epsilon}\vb{\cdot}\fd{\vb{\cal P}}{\bf E} + \pd{\bf B}{\epsilon}\vb{\cdot}\fd{\vb{\cal P}}{\bf B}\right),
\label{eq:Momentum_epsilon}
\end{equation}
where
\begin{equation}
\pd{\bf E}{\epsilon}\bdot\fd{\vb{\cal P}}{\bf E} + \pd{\bf B}{\epsilon}\bdot\fd{\vb{\cal P}}{\bf B} \;=\; \pd{\bf E}{\epsilon}\btimes\frac{\bf B}{4\pi c} + \frac{\bf E}{4\pi c}\btimes\pd{\bf B}{\epsilon},
\end{equation}
and
\begin{eqnarray} 
\frac{d{\bf p}}{d\epsilon} &=& \{ {\bf p},\; {\sf S}\} \;-\; \frac{e}{c}\,\pd{\bf A}{\epsilon} \nonumber \\
 &=& -\; \left(\nabla {\sf S} + \frac{e}{c}\,\pd{\bf A}{\epsilon} \right) \;+\; \frac{e}{c}\pd{\sf S}{\bf p}\btimes{\bf B}.
\end{eqnarray}
Using Eqs.~\eqref{eq:P_epsilon} and \eqref{eq:P_S}, we obtain
\begin{eqnarray}
\pd{\vb{\cal P}}{\epsilon} &=& -\;\int_{\bf z}{\sf f}\,\left(\nabla {\sf S} + \frac{e}{c}\,\pd{\bf A}{\epsilon} \right) \nonumber \\
 &&+ \int_{\bf x}\left(\pd{\mathbb{D}}{\epsilon}\btimes\frac{\bf B}{4\pi c} + \frac{\bf E}{4\pi c}\btimes\pd{\bf B}{\epsilon} \right).
\label{eq:Momentum_final}
\end{eqnarray}
In Eqs.~\eqref{eq:Energy_final} and \eqref{eq:Momentum_final}, polarization enters into the perturbative evolutions of the energy-momentum functionals explicitly through $\partial\mathbb{D}/\partial\epsilon$.

\subsection{Perturbed ideal MHD energy-momentum}

The perturbation derivative \eqref{eq:F_epsilon} of the ideal MHD energy functional is
\begin{eqnarray}
\pd{\cal H}{\epsilon} &=& \int_{\bf x} \left(\rho{\bf u}\bdot\pd{\bf u}{\epsilon} \;+\; \frac{\bf B}{4\pi}\bdot\pd{\bf B}{\epsilon} \right) \nonumber \\
 &&+\; \int_{\bf x} \vb{\xi}\bdot\left(\nabla P + \frac{1}{2}\,\rho\,\nabla|{\bf u}|^{2} \right)  \label{eq:delta_H_MHD} \\
  &=& \int \left[\rho{\bf u}\bdot\frac{d\vb{\xi}}{dt} + \vb{\xi}\bdot\nabla P + \frac{\bf B}{4\pi}\bdot\nabla\btimes(\vb{\xi}\btimes{\bf B}) \right],
\nonumber
\end{eqnarray}
where $\partial{\bf u}/\partial\epsilon$ is defined by Eq.~\eqref{eq:u_epsilon} and $\partial{\bf B}/\partial\epsilon = \nabla\btimes(\vb{\xi}\btimes{\bf B})$. Through a number of integrations by parts, Hameiri \cite{Hameiri_2003} has shown how Eq.~\eqref{eq:delta_H_MHD} can be expressed as 
\begin{eqnarray} 
\pd{\cal H}{\epsilon} &=& \int_{\bf x} \left[ \vb{\xi}\bdot\left(\rho\,{\bf u}\bdot\nabla{\bf u} + \nabla P - {\bf J}\btimes{\bf B}\right) + \nabla\bdot(\rho\,{\bf u})\;\fd{\cal S}{\rho} \right. \nonumber \\
 &&\left.+ {\bf u}\bdot\nabla s\;\fd{\cal S}{s} - \nabla\btimes({\bf u}\btimes{\bf B})\bdot\fd{\cal S}{\bf B} \right] \equiv -\,\pd{\cal S}{t},
 \end{eqnarray}
 which is again expected from the antisymmetry of the bracket $[\;,\;]$: $\partial{\cal H}/\partial\epsilon = -\,[{\cal S}, {\cal H}] \equiv -\,\partial{\cal S}/\partial t$.

The perturbation derivative \eqref{eq:F_epsilon} of the ideal MHD momentum functional, on the other hand, is
\begin{eqnarray}
\pd{\vb{\cal P}}{\epsilon} & = & \int_{\bf x} \left[ \rho \left(\frac{d\vb{\xi}}{dt} - \vb{\xi}\bdot\nabla{\bf u}\right) - {\bf u}\;\nabla\bdot(\rho\,\vb{\xi}) \right] \nonumber \\
 &=& \int_{\bf x}\left[ \rho\,\frac{d\vb{\xi}}{dt} \;-\; \nabla\bdot\left(\vb{\xi}\;\rho{\bf u}\right) \right] = \int_{\bf x}\rho\,\frac{d\vb{\xi}}{dt}.
 \end{eqnarray}
 
 In the next subsection, we will explore how the second-order perturbation derivative of the Hamiltonian functionals can be used to investigate plasma stability.

\subsection{Stability analyses and dynamical accessibility}

The stability analyses of Vlasov-Maxwell and ideal MHD equilibria have a long and rich history in plasma physics \cite{BFKK_1958,Kruskal_Oberman_1958,Frieman_Rotenberg_1960}. Here, only dynamically-accessible perturbation variations are considered \cite{PJM_1998} since they automatically preserve all Casimir invariants: $\delta{\cal C} = (\partial{\cal C}/\partial\epsilon)_{\epsilon = 0} \equiv 0$, since $[{\cal C}, {\cal G}] \equiv 0$ for all functionals ${\cal G}$. 

For the purpose of investigating dynamically-accessible plasma stability, we consider the second-order variation of the energy functional
\begin{equation}
\delta^{2}{\cal H} \;\equiv\; \frac{1}{2}\;\left.\frac{\partial^{2}{\cal H}}{\partial\epsilon^{2}}\right|_{\epsilon = 0}
\label{eq:delta_2_H}
\end{equation}
derived from the second-order perturbation derivative of the energy functional: 
\begin{eqnarray}
\frac{\partial^{2}{\cal H}}{\partial\epsilon^{2}} &=& \left[\pd{\cal H}{\epsilon},\;{\cal S}\right] = \int_{\bf x}\pd{\psi^{a}}{\epsilon}\;\fd{}{\psi^{a}}\left(\pd{\cal H}{\epsilon}\right) \nonumber \\
 &\equiv& \left[ \left[{\cal H},\; {\cal S}\right],\frac{}{} {\cal S} \right].
\label{eq:partial_2_H}
\end{eqnarray} 

\subsubsection{Vlasov-Maxwell stability}

We first consider the Vlasov-Maxwell case, where, using the first variation
\[ \delta\left(\pd{\cal H}{\epsilon}\right) = \int_{\bf z} \delta {\sf f}\;\frac{d{\sf K}}{d\epsilon} \;+\;  \int_{\bf x} \frac{1}{4\pi}\left(\delta{\bf E}\bdot\pd{\bf E}{\epsilon} + \delta{\bf B}\bdot\pd{\bf B}{\epsilon}\right), \]
Eq.~\eqref{eq:partial_2_H} becomes
\begin{equation}
\frac{\partial^{2}{\cal H}}{\partial\epsilon^{2}} = \int_{\bf z} {\sf f}\;\frac{d^{2}{\sf K}}{d\epsilon^{2}} \;+\; \frac{1}{4\pi} \int_{\bf x} \left( \left|\pd{\bf E}{\epsilon}\right|^{2}\;+\; \left|\pd{\bf B}{\epsilon}\right|^{2} \right),
\end{equation}
where
\[ \frac{d^{2}K}{d\epsilon^{2}} \;=\; \left\{ \frac{d{\sf K}}{d\epsilon},\; {\sf S} \right\} \;-\; \frac{e}{c}\pd{\bf A}{\epsilon}\bdot\pd{}{\bf p}\left(\frac{d{\sf K}}{d\epsilon}\right). \]
We can also integrate by parts the kinetic term to obtain
\begin{eqnarray*} 
\int_{\bf z} {\sf f}\;\frac{d^{2}{\sf K}}{d\epsilon^{2}} &=& -\;\int_{\bf z} \frac{d{\sf K}}{d\epsilon}\left( \left\{ {\sf f},\frac{}{} {\sf S}\right\} - \frac{e}{c}\pd{\bf A}{\epsilon}\bdot\pd{\sf f}{\bf p}\right) \\
 &\equiv& \int_{\bf z}\pd{\sf f}{\epsilon}\;\frac{d{\sf K}}{d\epsilon}.
 \end{eqnarray*}
 Hence, the second variation of the Vlasov-Maxwell energy functional is expressed as
 \begin{equation}
 \delta^{2}{\cal H} = -\,\frac{1}{2} \int_{\bf z}\left(\frac{d{\sf K}}{d\epsilon}\right)_{\epsilon = 0}^{2}{\sf f}_{0}^{\prime}({\sf K}) + \frac{1}{8\pi} \int_{\bf x} \left( |\delta{\bf E}|^{2} + |\delta{\bf B}|^{2} \right),
\end{equation}
which is identical to the second-order Vlasov-Maxwell free energy obtained by Morrison and Pfirsch \cite{PJM_DP_1989}. Here, the unperturbed Vlasov distribution ${\sf f}_{0}({\sf K})$ is assumed to be a function of the kinetic energy only and a sufficient condition for stability is ${\sf f}_{0}^{\prime}({\sf K}) < 0$.

\subsubsection{Ideal MHD stability}

For the ideal MHD case, we begin with the variation of the first-order perturbation derivative of the ideal MHD energy functional \eqref{eq:delta_H_MHD}:
\begin{eqnarray*}
\delta\left(\pd{\cal H}{\epsilon}\right) &=& \int_{\bf x}\left[ \delta\rho\;{\bf u}\bdot\frac{d\vb{\xi}}{dt} \;+\; \delta{\bf u}\bdot\rho\left(\frac{d\vb{\xi}}{dt} + \nabla\vb{\xi}\bdot{\bf u}\right) \right] \\
 &&+\; \int_{\bf x}\vb{\xi}\bdot\nabla\left(\delta\rho\;\pd{P}{\rho} \;+\; \delta s\;\pd{P}{s}\right) \\
  &&+\; \int_{\bf x}\left[\frac{\delta{\bf B}}{4\pi}\bdot\pd{\bf B}{\epsilon} \;+\; \frac{\bf B}{4\pi}\bdot\nabla\btimes\left(\vb{\xi}\btimes\delta{\bf B}\right) \right]
\end{eqnarray*}
which yields, after integration by parts, the following functional derivatives
\begin{eqnarray}
\fd{}{\rho}\left(\pd{\cal H}{\epsilon}\right) & = & {\bf u}\bdot\frac{d\vb{\xi}}{dt} \;-\; (\nabla\bdot\vb{\xi})\;\pd{P}{\rho}, \\
\fd{}{\bf u}\left(\pd{\cal H}{\epsilon}\right) & = & \rho \left( \frac{d\vb{\xi}}{dt}  \;+\; \nabla\vb{\xi}\bdot{\bf u}\right), \\
\fd{}{s}\left(\pd{\cal H}{\epsilon}\right) & = & -\;(\nabla\bdot\vb{\xi})\;\pd{P}{s}, \\
\fd{}{\bf B}\left(\pd{\cal H}{\epsilon}\right) & = & \frac{1}{4\pi} \left( \pd{\bf B}{\epsilon} \;+\; (\nabla\btimes{\bf B})\btimes\vb{\xi} \right).
\end{eqnarray}
After additional integration by parts, and using Eq.~\eqref{eq:u_epsilon}, the second-order perturbation derivative of the energy functional is expressed as $\partial^{2}{\cal H}/\partial\epsilon^{2} \equiv \partial^{2}{\cal K}/\partial\epsilon^{2} + \partial^{2}{\cal W}/\partial\epsilon^{2}$, where we have divided the energy functional ${\cal H}$ into the kinetic energy functional ${\cal K} \equiv \frac{1}{2}\int_{\bf x}\rho\,|{\bf u}|^{2}$, with $\partial{\cal K}/\partial\epsilon = \int_{\bf x}\rho{\bf u}\bdot d\vb{\xi}/dt$, and
\begin{equation}
\frac{\partial^{2}{\cal K}}{\partial\epsilon^{2}} = \int_{\bf x} \left[ \rho\;\pd{\bf u}{\epsilon}\vb{\cdot}\left(\frac{d\vb{\xi}}{dt} + \nabla\vb{\xi}\vb{\cdot}{\bf u}\right) + \pd{\rho}{\epsilon}\;{\bf u}\bdot\frac{d\vb{\xi}}{dt} \right], 
 \end{equation}
 and the potential energy functional ${\cal W} \equiv {\cal H} - {\cal K}$, with $\partial{\cal W}/\partial\epsilon = \int_{\bf x} \left[ \vb{\xi}\bdot\nabla P + \nabla\btimes(\vb{\xi}\btimes{\bf B})\bdot{\bf B}/4\pi \right]$, and
\begin{eqnarray*}
\frac{\partial^{2}{\cal W}}{\partial\epsilon^{2}} &=&  \int_{\bf x} \left[ (\nabla\bdot\vb{\xi})\;\vb{\xi}\bdot\nabla P \;+\; \rho\,\pd{P}{\rho}\;(\nabla\bdot\vb{\xi})^{2} \right] \\
 &&+ \frac{1}{4\pi} \int_{\bf x}\left( \left|\pd{\bf B}{\epsilon}\right|^{2} + (\nabla\btimes{\bf B})\btimes\vb{\xi}\bdot\pd{\bf B}{\epsilon}\right).
\end{eqnarray*}
We therefore obtain the standard ideal MHD energy principle $\delta^{2}{\cal W} \equiv \frac{1}{2}\,(\partial^{2}{\cal W}/\partial\epsilon^{2})_{\epsilon = 0}$ \cite{BFKK_1958,Frieman_Rotenberg_1960}:
\begin{eqnarray}
\delta^{2}{\cal W} & = & \frac{1}{2} \int_{\bf x} \left[ (\nabla\bdot\vb{\xi})\;\vb{\xi}\bdot\nabla P_{0} \;+\frac{}{} \gamma\,P_{0}\;(\nabla\bdot\vb{\xi})^{2} \right. \nonumber \\
 &&\left.+\; \frac{1}{4\pi} \left( \left|{\bf B}_{1}\right|^{2} \;+\; (\nabla\btimes{\bf B}_{0})\btimes\vb{\xi}\bdot{\bf B}_{1}\right) \right],
 \end{eqnarray}
 where ${\bf B}_{1} = \nabla\btimes(\vb{\xi}\btimes{\bf B}_{0})$ and $\gamma\,P_{0} \equiv \rho_{0}\,P^{\prime}_{0}(\rho_{0})$.
 
 We note that higher order perturbative derivatives can be considered to investigate marginal stability (in which $\delta^{2}{\cal W} = 0$). Indeed, a cubic (third-order) energy principle $\delta^{3}{\cal W} \equiv \frac{1}{6}\,(\partial^{3}{\cal W}/\partial\epsilon^{3})_{\epsilon = 0}$ can be derived from the third-order perturbative derivative
 \begin{eqnarray}
\frac{\partial^{3}{\cal W}}{\partial\epsilon^{3}} &=& \int_{\bf x}\left[\pd{\rho}{\epsilon}\fd{}{\rho}\left(\frac{\partial^{2}{\cal W}}{\partial\epsilon^{2}} \right) + \pd{s}{\epsilon}\fd{}{s}\left(\frac{\partial^{2}{\cal W}}{\partial\epsilon^{2}} \right) \right. \nonumber \\
 &&\left.+\; \pd{\bf B}{\epsilon}\bdot\fd{}{\bf B}\left(\frac{\partial^{2}{\cal W}}{\partial\epsilon^{2}} \right) \right],
\end{eqnarray}
where
\begin{widetext}
\begin{eqnarray*}
\delta\left(\frac{\partial^{2}{\cal W}}{\partial\epsilon^{2}}\right) &=&   \int_{\bf x} \left[ (\nabla\bdot\vb{\xi})\;\vb{\xi}\bdot\nabla\left(\delta\rho\; \pd{P}{\rho} + \delta s\;\pd{P}{s}\right) \;+\; \delta\rho\,\pd{}{\rho}\left(\rho\,\pd{P}{\rho}\right)\;(\nabla\bdot\vb{\xi})^{2} 
+ \rho\,\delta s\,\frac{\partial^{2}P}{\partial s\partial\rho}\,(\nabla\bdot\vb{\xi})^{2}\right] \\
 &&+ \frac{1}{4\pi} \int_{\bf x} \left\{ \pd{\bf B}{\epsilon}\bdot\left[ 2\,\nabla\btimes\left(\vb{\xi}\btimes\delta{\bf B}\right) \;+\frac{}{} (\nabla\btimes\delta{\bf B})\btimes\vb{\xi} \right] + (\nabla\btimes{\bf B})\btimes\vb{\xi}\bdot\nabla\btimes\left(\vb{\xi}\btimes\delta{\bf B}\right) \right\},
 \end{eqnarray*}
\end{widetext} 
from which functional derivatives $\delta(\partial_{\epsilon}^{2}{\cal W})/\delta\psi^{a}$ can be calculated. Similar expressions have been considered in the context of marginal ideal MHD stability by Pfirsch and Sudan \cite{Pfirsch_Sudan_1993} and the derivation of Manley-Rowe coupling coefficients for nonlinear three-wave ideal MHD interactions by Hirota \cite{Hirota_2011}. In both cases, the third-order functionals are cubic expressions in powers of the fluid displacement $\vb{\xi}$. The third-order perturbation Lagrangian for the perturbed Vlasov-Maxwell equations has been derived in Ref.~\cite{Brizard_2018}, which could also be used to investigate higher-order plasma stability and resonant nonlinear three-wave interactions.

 \section{\label{sec:summary}Summary and Prospects}
 
 In the present work, the Hamiltonian formulations of the perturbative Vlasov-Maxwell equations and the perturbative ideal magnetohydrodynamics are given in terms of a theoretical functional representation. In each representation, the reduced polarization and magnetization \eqref{eq:P_epsilon} and \eqref{eq:M_epsilon} not only play a crucial role in perturbative Vlasov-Maxwell theory, but also in the Clebsch representation \eqref{eq:EB_MHD} of ideal magnetohydrodynamics.
 
The central role of polarization in Vlasov perturbation theory is perhaps not surprising, because Vlasov perturbations can only involve displacements of infinitesimal elements in phase space that conserve particle numbers. Hence, phase-space displacements $d{\bf x}/d\epsilon$ that are species-dependent (in a quasi-neutral plasma environment) naturally lead to finite polarization \eqref{eq:P_epsilon}. What is perhaps surprising is that, in the ideal MHD variational principle based on the ideal MHD action functional \eqref{eq:Lag_MHD}, the variation with respect to polarization leads to the ideal MHD constraint \eqref{eq:P_var} when the generalized Clebsch representation \eqref{eq:EB_MHD} is used. 

In conclusion, we note that several Hamiltonian representations have been found for the Hall and Extended MHD equations \cite{Hirota_2006,DAvignon_2016,Kaltsas_2020} as well as various reduced plasma-fluid models (e.g., reduced ideal MHD equations \cite{Morrison_Hazeltine_1984} and gyrofluid equations \cite{Tassi_2019,Tronci_2020}), which are now amenable to Hamiltonian perturbation theory as represented in this paper. The Hamiltonian perturbation framework presented here can also be applied to the Hamiltonian structures of kinetic-MHD equations and reduced plasma models (e.g., the gyrokinetic Vlasov-Maxwell equations). These reduced Vlasov-Maxwell equations will be of particular interest in future work since they involve perturbed Vlasov-Maxwell brackets \cite{Brizard_2016}, with perturbation polarization and magnetization already imbedded in them.
  
\begin{acknowledgments}
The Authors wish to thank Dr.~Emanuele Tassi for reminding them about the connections between our work and the works of Morrison and his colleagues on dynamical accessibility.  We are also grateful to Prof.~Phil Morrison for making useful suggestions in improving the manuscript in terms of its historical accuracy. AJB acknowledges support from a U.S. DoE grant under contract DE-SC0014032 and an NSF grant under contract PHY-1805164. This work has been carried out within the framework of the French Federation for Magnetic Fusion Studies (FR-FCM) and of the Eurofusion consortium, and has received funding from the Euratom research and training programme 2014-2018 and 2019-2020 under grant agreement No 633053. The views and opinions expressed herein do not necessarily reflect those of the European Commission. 
\end{acknowledgments}

\vspace*{0.1in}

\bc
{\bf Data Availability Statement}
\ec

Data sharing is not applicable to this article as no new data were created or analyzed in this study.

\appendix

\section{\label{sec:App_A}Perturbation Variational Principles}

\subsection{Perturbed single-particle dynamics}

In this Appendix, we consider the least-action principle for single-particle dynamics $\delta A[{\cal C}] = 0$ expressed in terms of the action integral \cite{Brizard_2001}
\begin{equation}
A[{\cal C}] \equiv \int_{\cal C}\gamma = \int_{\cal C} \left[\left(\frac{e}{c}\,{\bf A} + {\bf p}\right)\bdot\exd{\bf x} - {\sf H}\;\exd t - {\sf S}\,\exd\epsilon\right],
\end{equation}
which is defined along an open path ${\cal C}$ with fixed end points in the parameter space $(t,\epsilon)$. Stationarity of the action integral with respect to arbitrary phase-space variations $(\delta{\bf x}, \delta{\bf p})$, which vanish at the end points of the open path ${\cal C}$, yields the Euler-Lagrange equations
\begin{eqnarray}
d{\bf x} & = & \pd{{\sf H}}{\bf p}\;dt \;+\; \pd{\sf S}{\bf p}\;d\epsilon, \label{eq:dx} \\
d{\bf p} & = & e\,{\bf E}\;dt + \frac{e}{c}\,d{\bf x}\btimes{\bf B} - \left(\nabla {\sf S} + \frac{e}{c}\pd{\bf A}{\epsilon}\right) d\epsilon, \label{eq:dp}
\end{eqnarray}
from which we recover Eqs.~\eqref{eq:Ham_h}-\eqref{eq:Ham_S}.

We now require that the equations \eqref{eq:dx}-\eqref{eq:dp} be valid for any open path ${\cal C}$ with the same fixed end points. For this purpose, we consider the integral along the closed loop $\partial{\cal A} \equiv {\cal C} - {\cal C}'$, where the open surface 
${\cal A}$ in the parameter space $(t,\epsilon)$ denotes the area enclosed by $\partial{\cal A}$. Using Stokes' Theorem,  we obtain 
\[ \oint_{\partial{\cal A}}\gamma \;=\; \int_{\cal A}\,\exd\gamma \;\equiv\; \int_{\cal A}\Lambda_{\epsilon}\;\exd\epsilon\wedge\exd t, \]
where the Lagrangian scalar field
\begin{eqnarray}
\Lambda_{\epsilon} & = & \frac{d{\bf x}}{d\epsilon}\bdot\left( e{\bf E} \;+\; \frac{e}{c}\frac{d{\bf x}}{dt}\btimes{\bf B} \;-\; \frac{d{\bf p}}{dt}\right) \nonumber \\ 
 & &+ \frac{d{\bf p}}{d\epsilon}\bdot\left( \frac{d{\bf x}}{dt} - \frac{\bf p}{m}\right) + \frac{d{\sf S}}{dt} - e\,\left( \pd{\Phi}{\epsilon} - \frac{1}{c}\frac{d{\bf x}}{dt}\bdot\pd{\bf A}{\epsilon}\right)
 \nonumber \\
  & \equiv & \frac{d{\sf S}}{dt} - e\,\left( \pd{\Phi}{\epsilon} - \frac{1}{c}\frac{d{\bf x}}{dt}\bdot\pd{\bf A}{\epsilon}\right)
   \label{eq:Lambda}
 \end{eqnarray}
is defined after making use of Eqs.~\eqref{eq:dx}-\eqref{eq:dp}. The condition of path independence, therefore, requires that $\Lambda_{\epsilon} \equiv 0$, from which we recover Eq.~\eqref{eq:Sh_constraint}.
 
 Lastly, we rewrite Eq.~\eqref{eq:Lambda} as
\begin{eqnarray}
\Lambda_{\epsilon}  & = & \pd{\sf S}{t} \;-\; \pd{{\sf H}}{\epsilon} \;+\; \pd{{\sf H}}{\bf p}\bdot\left(\nabla{\sf S} \;+\; \frac{e}{c}\,\pd{\bf A}{\epsilon}\right) \nonumber \\
    &&+\; e \left({\bf E} \;+\; \frac{\bf v}{c}\btimes{\bf B}\right)\bdot\pd{\sf S}{\bf p},
 \end{eqnarray}
 so that the partial derivatives
 \begin{equation}
 \left(\pd{\Lambda_{\epsilon}}{\bf E},\; \pd{\Lambda_{\epsilon}}{\bf B}\right) \;=\; \left( e\,\frac{d{\bf x}}{d\epsilon},\frac{e}{c}\,\frac{d{\bf x}}{d\epsilon}\btimes\frac{d{\bf x}}{dt}\right),
 \end{equation}
 can be used to define the polarization and magnetization
 \begin{equation}
 \left(\mathbb{P}_{\epsilon},\; \mathbb{M}_{\epsilon}\right) \;\equiv\; \int_{\bf p} \left(\pd{\Lambda_{\epsilon}}{\bf E},\; \pd{\Lambda_{\epsilon}}{\bf B}\right)\;{\sf f},
 \label{eq:Lambda_EB}
 \end{equation} 
 which appear in Eqs.~\eqref{eq:P_epsilon} and \eqref{eq:M_epsilon}.

\subsection{Variational Principle for Perturbed Vlasov-Maxwell Equations}

We now present the variational principle $\delta\int {\cal L}_{\epsilon}\,dt = 0$ from which the perturbed Vlasov-Maxwell equations are derived. Here, the perturbation Lagrangian functional \cite{Brizard_2018} is
\begin{equation}
{\cal L}_{\epsilon} \;=\; \int_{\bf z} {\sf f}\;\Lambda_{\epsilon} \;+\; \int_{\bf x} \frac{1}{4\pi} \left({\bf E}\bdot\pd{\bf E}{\epsilon} \;-\; {\bf B}\bdot\pd{\bf B}{\epsilon}\right),
\label{eq:caL_e}
\end{equation}
where  the variational fields are $(f,{\bf E},{\bf B})$ and the Lagrangian scalar field is defined as
\begin{eqnarray} 
\Lambda_{\epsilon} & = & \left(\pd{\sf S}{t} \;-\; e\,\pd{\Phi}{\epsilon}\right) \;+\; {\bf v}\bdot\left(\nabla{\sf S} \;+\; \frac{e}{c}\,\pd{\bf A}{\epsilon}\right) \nonumber \\
 &&+\; e\,\left({\bf E} + \frac{\bf v}{c}\btimes{\bf B}\right)\bdot\pd{\sf S}{\bf p}.
 \end{eqnarray}
The perturbation Lagrangian functional \eqref{eq:caL_e} was recently used \cite{Brizard_2018} to explore the perturbation variational structure of the Vlasov-Maxwell equations.

 The variations of the perturbation Lagrangian functional \eqref{eq:caL_e}:
  \begin{eqnarray}
\delta{\cal L}_{\epsilon} & = & \int_{\bf z} \left[\delta {\sf f}\;\Lambda_{\epsilon} + {\sf f}\frac{}{}\left( \delta{\bf E}\bdot\pd{\Lambda_{\epsilon}}{\bf E} + \delta{\bf B}\bdot\pd{\Lambda_{\epsilon}}{\bf B} \right) \right] \nonumber \\
 &&+\; \int_{\bf x} \frac{1}{4\pi} \left(\delta{\bf E}\bdot\pd{\bf E}{\epsilon} \;-\; \delta{\bf B}\bdot\pd{\bf B}{\epsilon}\right),
\label{eq:delta_caL_e}
\end{eqnarray} 
yield the identities
 \begin{eqnarray}
 \fd{{\cal L}_{\epsilon}}{\sf f} & = & \Lambda_{\epsilon} \;\equiv\; 0, \\
 \fd{{\cal L}_{\epsilon}}{\bf E} & = & \int_{\bf p} e\,{\sf f}\;\frac{d{\bf x}}{d\epsilon} \;+\; \frac{1}{4\pi}\;\pd{\bf E}{\epsilon} \nonumber \\
   & = & c\,\nabla\btimes\left(\fd{\cal S}{\bf B}\right) \;\equiv\; \frac{1}{4\pi}\;\pd{\mathbb{D}}{\epsilon}, \\
 \fd{{\cal L}_{\epsilon}}{\bf B} & = & \int_{\bf p} e\,{\sf f}\;\left(\frac{d{\bf x}}{d\epsilon}\btimes\frac{d{\bf x}}{dt}\right) \;-\; \frac{1}{4\pi}\;\pd{\bf B}{\epsilon} \nonumber \\
   & = & -\; \pd{}{t}\left( \fd{\cal S}{\bf B}\right) \;\equiv\; -\;\frac{1}{4\pi}\;\pd{\mathbb{H}}{\epsilon},
 \end{eqnarray}
 where we used Eqs.~\eqref{eq:P_S}-\eqref{eq:M_S}, respectively, with Eq.~\eqref{eq:Lambda_EB} and $d{\bf x}/d\epsilon \equiv \partial {\sf S}/\partial{\bf p}$ and $d{\bf x}/dt = \partial {\sf H}/\partial{\bf p} = {\bf v}$. Hence, by combining these results, the Eulerian variation of the perturbation 
 Lagrangian functional \eqref{eq:caL_e} yields the expression
 \begin{eqnarray}
 \delta{\cal L}_{\epsilon} & = & \int_{\bf z}\delta {\sf f}\;\fd{{\cal L}_{\epsilon}}{\sf f} \;+\; \int_{\bf x} \left(\delta{\bf E}\bdot\fd{{\cal L}_{\epsilon}}{\bf E} \;+\; \delta{\bf B}\bdot\fd{{\cal L}_{\epsilon}}{\bf B} \right) \nonumber \\
  & = & \int_{\bf x} \frac{1}{4\pi} \left( \delta{\bf E}\bdot \pd{\mathbb{D}}{\epsilon} \;-\; \delta{\bf B}\bdot \pd{\mathbb{H}}{\epsilon}\right) \nonumber \\
   & = & -\;\pd{}{t}\left( \int_{\bf x}\delta{\bf B}\bdot\fd{\cal S}{\bf B} \right),
  \end{eqnarray}
where we assumed that the field variations $(\delta{\bf E},\delta{\bf B})$ satisfy the Faraday constraint $c\nabla\btimes\delta{\bf E} + \partial\delta{\bf B}/\partial t = 0$. 

\section{\label{sec:App_B}Hamiltonian Constraint}

In this Appendix, we derive the Hamiltonian constraint \eqref{eq:constraint}. We greatly simplify the derivation by adopting an extended phase-space representation whereby the Hamiltonian functions $({\sf H},{\sf S})$ are extended to $({\sf H}^{*} \equiv {\sf H} - \eta, {\sf S}^{*} \equiv {\sf S} - \zeta)$, and the extended Poisson bracket is
\begin{widetext}
\begin{eqnarray}
\{ F,\;G\}^{*} &\equiv& \{ F,\; G\} \;+\; \frac{e}{c} \pd{F}{\bf p}\bdot \left( \pd{\bf A}{t}\,\pd{G}{\eta} \;+\; \pd{\bf A}{\epsilon}\,\pd{G}{\zeta} \right) \;-\; \frac{e}{c} \pd{G}{\bf p}\bdot \left( \pd{\bf A}{t}\,\pd{F}{\eta} \;+\; \pd{\bf A}{\epsilon}\,\pd{F}{\zeta} \right) 
\nonumber \\
 &&+\; \left(\pd{F}{\eta}\,\pd{G}{t} - \pd{F}{t}\,\pd{G}{\eta}\right) \;+\;  \left(\pd{F}{\zeta}\,\pd{G}{\epsilon} - \pd{F}{\epsilon}\,\pd{G}{\zeta}\right),
 \label{eq:PB_ext}
\end{eqnarray}
\end{widetext}
which is obtained by adding the vector-potential terms $(\partial{\bf A}/\partial t, \partial{\bf A}/\partial\epsilon)$ and the canonical-pair terms associated with time $(t,\eta)$ and perturbation $(\epsilon,\zeta)$ to the standard Poisson bracket \eqref{eq:xv_PB}. Hence, the operators \eqref{eq:d_dt}-\eqref{eq:d_de} become $d/dt = \{\;, {\sf H}^{*}\}^{*}$ and $d/d\epsilon = \{\;, {\sf S}^{*}\}^{*}$. We note that the extended Poisson bracket \eqref{eq:PB_ext} is obtained from the extended symplectic one-form 
\[ \gamma^{*} \;=\; (e{\bf A}/c + {\bf p})\bdot\exd{\bf x} - \eta\,\exd t - \zeta\,\exd\epsilon \]
by inversion of the Lagrange two-form $\omega^{*} \equiv \exd\gamma^{*}$. Since the two-form $\omega^{*}$ is closed (i.e., $\exd\omega^{*} = \exd^{2}\gamma^{*} = 0$, which requires 
$\nabla\bdot{\bf B} = 0$),  the extended Poisson bracket \eqref{eq:PB_ext} automatically satisfies the Jacobi property.

Our derivation of the Hamiltonian constraint \eqref{eq:constraint} now proceeds simply from the fact that the extended Poisson bracket satisfies the extended version of the Jacobi identity. First, we write
\begin{widetext}
\[ \frac{d}{dt}\left(\frac{df}{d\epsilon}\right) - \frac{d}{d\epsilon}\left(\frac{df}{dt}\right) \;=\; \left\{ \{ f,\; {\sf S}^{*}\}^{*},\frac{}{} {\sf H}^{*}\right\}^{*} -  \left\{ \{ f,\; {\sf H}^{*}\}^{*},\frac{}{} {\sf S}^{*}\right\}^{*} \;=\; \left\{ \{ f,\; {\sf S}^{*}\}^{*},\frac{}{} {\sf H}^{*}\right\}^{*} + 
\left\{ \{ {\sf H}^{*}, f\}^{*},\frac{}{} {\sf S}^{*}\right\}^{*}, \]
\end{widetext}
where we used the antisymmetry of the extended Poisson bracket \eqref{eq:PB_ext} in the second term on the right. We now use the Jacobi identity to obtain the commutation relation
\begin{equation} 
0 \equiv \frac{d}{dt}\left(\frac{df}{d\epsilon}\right) - \frac{d}{d\epsilon}\left(\frac{df}{dt}\right) = -\;\left\{ \{ {\sf S}^{*},\; {\sf H}^{*}\}^{*},\frac{}{} f\right\}^{*},
\end{equation}
which must be valid for any function $f$. Therefore, this commutation relation implies that
\begin{eqnarray}
0 = \left\{ {\sf S}^{*},\; {\sf H}^{*} \right\}^{*} &=& \left(\pd{\sf S}{t} - \pd{\sf H}{\epsilon}\right) +  \{ {\sf S},\; {\sf H}\} \nonumber \\
 &&-\; \frac{e}{c} \left(\pd{\sf S}{\bf p}\bdot\pd{\bf A}{t} - \pd{\sf H}{\bf p}\bdot\pd{\bf A}{\epsilon}\right) \nonumber \\
 &\equiv& \frac{d{\sf S}}{dt} \;-\; \frac{d{\sf H}}{d\epsilon} \;-\; \{{\sf S},\; {\sf H}\}, 
 \end{eqnarray}
 and, thus, the Hamiltonian constraint \eqref{eq:constraint} is recovered.


\begin{thebibliography}{99}

\bibitem{Davidson} R.~C.~Davidson, {\it Methods in Nonlinear Plasma Theory}, Academic Press (1972).

\bibitem{Cary_Kaufman_1981} J.~R.~Cary and A.~N.~Kaufman, Phys.~Fluids {\bf 24}, 1238 (1981).

\bibitem{Dewar_1976} R.~L.~Dewar, J.~Phys.~A: Math.~Gen.~{\bf 9}, 2043 (1976).

\bibitem{Kaufman_1978} A.~N.~Kaufman, {\it The Lie transform: A new approach to classical perturbation theory}, in AIP Conference Proceedings {\bf 46}, 286 (1978).

\bibitem{Cary_Brizard_2009} J.~R.~Cary and A.~J.~Brizard, Rev.~Mod.~Phys.~{\bf 81}, 693 (2009).

\bibitem{Brizard_Hahm_2007} A.~J.~Brizard and T.~S.~Hahm, Rev.~Mod.~Phys.~{\bf 79}, 421 (2007).

\bibitem{Brizard_2001} A.~J.~Brizard, Phys.~Lett.~A {\bf 291}, 146 (2001).

\bibitem{Brizard_2018} A.~J.~Brizard, Phys.~Plasmas {\bf 25}, 112112 (2018).

\bibitem{Brizard_2008} A.~J.~Brizard, Comm.~Nonlin.~Sci.~Num.~Sim.~{\bf 13}, 24 (2008).

\bibitem{Brizard_2009} A.~J.~Brizard, J.~Phys.~Conf.~{\bf 169}, 012003 (2009).

\bibitem{Boyd_Turner_1978} T.J.M. Boyd and J.G. Turner, J.~Math.~Phys.~{\bf 19}, 1403 (1978).

\bibitem{Viscondi_2016} T.~F.~Viscondi, I.~L.~Caldas, and P.~J.~Morrison, J. Phys. A {\bf 49}, 165501 (2016).

\bibitem{Morrison_Greene_1980} P.~J.~Morrison and J.~M.~Greene, Phys.~Rev.~Lett.~{\bf 45}, 790 (1980); errata, Phys.~Rev.~Lett.~{\bf 48}, 569 (1982).

\bibitem{M} P.~J.~Morrison, Phys. Lett.~{\bf A80}, 383 (1980).

\bibitem{PJM_1982} P.~J.~Morrison, {\it Poisson brackets for fluids and plasmas}, AIP Conf.~Proc.~{\bf 88}, 13 (1982).

\bibitem{MW} J.~E.~Marsden, A.~Weinstein, Physica {\bf 4D}, 394 (1982).

\bibitem{B} I.~Bialynicki-Birula, J.~C.~Hubbard, L.~A.~Turski, Physica {\bf 128A}, 509 (1984).

\bibitem{Hameiri_2003} E.~Hameiri, Phys.~Plasmas {\bf 10}, 2643 (2003).

\bibitem{PJM_DP_1989} P.~J.~Morrison and D.~Pfirsch, Phys.~Rev.~A {\bf 40}, 3898 (1989).

\bibitem{PJM_DP_1990} P.~J.~Morrison and D.~Pfirsch, Phys.~Fluids B {\bf 2}, 1105 (1990).

\bibitem{Brizard_1994} A.~Brizard, Phys.~Plasmas {\bf 1}, 2473 (1994).

\bibitem{Hameiri_1998} E.~Hameiri, Phys.~Plasmas {\bf 5}, 3270 (1998).

\bibitem{Andreussi_2013} T.~Andreussi, P.~J.~Morrison, and F.~Pegoraro, Phys.~Plasmas {\bf 20}, 092104 (2013).

\bibitem{Ilgisonis_Pastukhov_2000} V.~I.~Ilgisonis and V.~P.~Pastukhov, JETP Lett.~{\bf 72}, 530 (2000).

\bibitem{PJM_1998} P.~J.~Morrison, Rev.~Mod.~Phys.~{\bf 70}, 467 (1998).

\bibitem{PJM_2005} P.~J.~Morrison, Phys.~Plasmas {\bf 12}, 058102 (2005).

\bibitem{Brizard_ERT_2003} A.~J.~Brizard and E.~R.~Tracy, Phys.~Plasmas {\bf 10}, 2163 (2003).

\bibitem{PJM_2013} P.~J.~Morrison, Phys.~Plasmas {\bf 20}, 012104 (2013).

\bibitem{Brizard_2016_arxiv} See App.~B in A.~J.~Brizard,  P.~J.~Morrison, J.~W.~Burby, L.~de Guillebon, and M.~Vittot, {\it Lifting of the Vlasov-Maxwell Bracket by Lie-transform Method}, arXiv:1606.06652 (2016). 

\bibitem{Littlejohn_1982} R.~G.~Littlejohn, J.~Math.~Phys.~{\bf 23}, 742 (1982).

\bibitem{Pfirsch_1966} D.~Pfirsch, Nuc.~Fusion {\bf 6}, 301 (1966).

\bibitem{Hori_1966} G.~Hori, Pub.~Astron.~Soc.~Japan {\bf 18}, 287 (1966).

\bibitem{Kawakami_1970} I.~Kawakami, J.~Phys.~Soc.~Japan {\bf 28}, 505 (1970).

\bibitem{Dewar_1973} R.~L.~Dewar, Phys.~Fluids {\bf 16}, 1102 (1973).

\bibitem{Brizard_2013} A.~J.~Brizard, Phys.~Plasmas {\bf 20}, 092309 (2013).

\bibitem{Healy_1978}  W.~P.~Healy, Proc.~R.~Soc.~Lond.~A {\bf 358}, 367 (1978).

\bibitem{Goedecke_1998} G.~H.~Goedecke, Am.~J.~Phys.~{\bf 66}, 1010 (1998).

\bibitem{Kaufman_Cohen_2019} A.~N.~Kaufman and B.~I.~Cohen, {\it Theoretical Plasma Physics}, J.~Plasma Phys.~{\bf 85}, 205850601 (2019).

\bibitem{Newcomb_1962} W.~A.~Newcomb, Nucl.~Fusion Suppl.~pt 2, 451 (1962).

\bibitem{Calkin_1963} M.~G.~Calkin, Can.~J.~Phys.~{\bf 41}, 2241 (1963).

\bibitem{Seliger_1968} R.~L.~Seliger and G.~B.~Whitham, Proc.~Roy.~Soc.~A {\bf 305}, 1 (1968).

\bibitem{Merches_1969} I.~Merches, Phys.~Fluids {\bf 12}, 2225 (1969).

\bibitem{Nassar_1985} A.~B.~Nassar and S.~J.~Putterman, Phys.~Fluids {\bf 28}, 1001 (1985).

\bibitem{Yahalom_2016} A.~Yahalom, J.~Plasma Phys.~{\bf 82}, 905820205 (2016).

\bibitem{Yahalom_Qin_2020} A.~Yahalom and H.~Qin, {\it Noether Currents for Eulerian Variational Principles in Non Barotropic Magnetohydrodynamics and Topological Conservations Laws}, arXiv:2005.14005 (2020).

\bibitem{footnote} The unusual sign in front of the Maxwell Lagrangian density in Eq.~\eqref{eq:Lag_MHD} is needed to ensure that the Legendre transformation $H = \sum_{a} \partial_{t}\chi^{a}\,\partial L/\partial(\partial_{t}\chi^{a}) - L$ produces the proper Hamiltonian density $H = \frac{1}{2}\,\rho\,|{\bf u}|^{2} + \rho\,U + (|{\bf E}|^{2} + |{\bf B}|^{2})/8\pi$ (up to an exact divergence).

\bibitem{Panofsky_Phillips_1955} W.~K.~H.~Panofsky and M.~Phillips, {\it Classical Electricity and Magnetism}, Addison-Wesley (1955), Sec.~9-4.

\bibitem{Roa-Neri_2018} J.-A.-E.~Roa-Neri, J.-L.~Jim\'{e}mez-Ram\'{i}rez, and I.~Campos-Flores, J.~Electromagnetic Analysis and Applications {\bf 10}, 185 (2018).

\bibitem{BFKK_1958} I.~B.~Bernstein, E.~A.~Frieman, M.~D.~Kruskal, and R.~M.~Kulsrud, Proc.~R.~Soc.~London, Ser. A {\bf 244}, 17 (1958).

\bibitem{Kruskal_Oberman_1958} M.~D.~Kruskal and C.~R.~Oberman, Phys.~Fluids {\bf 1}, 275 (1958).

\bibitem{Frieman_Rotenberg_1960} E.~Frieman and M.~Rotenberg, Rev.~Mod.~Phys.~{\bf 32}, 898 (1960).

\bibitem{Pfirsch_Sudan_1993} D.~Pfirsch and R.~N.~Sudan, Phys.~Fluids B {\bf 5}, 2052 (1993).

\bibitem{Hirota_2011} M.~Hirota, J.~Plasma Phys.~{\bf 77}, 589 (2011).

\bibitem{Hirota_2006} M.~Hirota, Z.~Yoshida, and E.~Hameiri, Phys.~Plasmas {\bf 13}, 022107 (2006).

\bibitem{DAvignon_2016} E.~C.~D'Avignon, P.~J.~Morrison, and M.~Lingam, Phys.~Plasmas {\bf 23}, 082101 (2016).

\bibitem{Kaltsas_2020} D.~A.~Kaltsas, G.~N.~Throumoulopoulos, and P.~J.~Morrison, Phys.~Plasmas {\bf 27}, 012104 (2020).

\bibitem{Morrison_Hazeltine_1984} P.~J.~Morrison and R.~D.~Hazeltine, Phys.~Fluids {\bf 27}, 886 (1984).

\bibitem{Tassi_2019} E.~Tassi, J.~Phys.~A: Math.~Theor.~{\bf 52} 465501 (2019).

\bibitem{Tronci_2020} C.~Tronci, Plasma Phys.~Control.~Fusion {\bf 62} 085006 (2020).

\bibitem{Brizard_2016} A.~J.~Brizard, P.~J.~Morrison, J.~W.~Burby, L.~de Guillebon, and M.~Vittot, J.~Plasma Phys.~{\bf 82}, 905820608 (2016).



\end{thebibliography}
\end{document}